\documentclass[useAMS,usenatbib]{mn2e}
\usepackage{graphics}
\usepackage[english]{babel}
\usepackage{amsmath,amssymb,amsfonts} 
\usepackage{anysize}                
\usepackage{color}                    
\usepackage{hyperref}                 
\usepackage{times}
\usepackage[none]{hyphenat}
\usepackage{graphicx}
\usepackage{latexsym}
\usepackage{babel}
\usepackage{longtable}
\usepackage{pdflscape}
\usepackage{fmtcount}

\newcommand{\Msun}{\ensuremath{{\rm M}_{\odot}}}

\newcommand{\bzk}{$BzK_{s}$ }
\newcommand{\gzk}{$gzK_{s}$ }

\newcommand{\gzhk}{$gzHK_{s}$ }
\newcommand{\zhk}{$zHK_{s}$ }
\newcommand{\Ks}{$K_{s}$ }
\newcommand{\zs}{$z\sim$ }
\newcommand{\s}{$\sim$ }
\newcommand{\ebv}{$E(B-V)$}
\newcommand{\zk}{$z - K_s$}
\newcommand{\Veff}{$V_{\mathrm{eff}}$}

\title[The numbers of star-forming and passive $z\sim2$ galaxies]{The numbers of \zs2 star-forming and passive galaxies in 2.5 square degrees of deep CFHT imaging}
\author[L.\ Arcila-Osejo and M.\ Sawicki]{Liz Arcila-Osejo\thanks{Email: osejo@ap.smu.ca, sawicki@ap.smu.ca} and Marcin Sawicki\\
Department of Astronomy $\&$ Physics, and the Institute for Computational Astrophysics, Saint Mary's University,\\ 923 Robie Street, Halifax, Nova Scotia, B3H 3C3, Canada}

\begin{document}

\date{Accepted 2013 July 19, for Publication in MNRAS}

\pagerange{\pageref{firstpage}--\pageref{lastpage}} \pubyear{2013}

\maketitle

\label{firstpage}

\begin{abstract}
We use an adaptation of the \bzk\ technique to select  $\sim$40,000 \zs2 galaxies (to $K_{AB}=24$), including $\sim$5,000 passively evolving (PE) objects (to $K_{AB}=23$), from 2.5 $deg^2$ of deep CFTH imaging. The passive galaxy luminosity function exhibits a clear peak at $R=22$ and a declining faint-end slope ($\alpha= -0.12 ^{+0.16}_{-0.14}$), while that of star-forming galaxies is characterized by a steep faint-end slope ($\alpha = -1.43\pm0.02(systematic)^{+0.05}_{-0.04}(random)$). The details of the LFs are somewhat sensitive (at $<$25\% level) to cosmic variance even in these large ($\sim$0.5 deg$^2$) fields, with the D2 field (located in the COSMOS field) most discrepant from the mean. The shape of the \zs2 stellar mass function of passive galaxies is remarkably similar to that at \zs0.9, save for a factor of $\sim$4 lower number density.  This similarity suggests that the same mechanism may be responsible for the formation of passive galaxies seen at both these epochs.  This same formation mechanism may also operate down to \zs0 if the local PE galaxy mass function, known to be two-component, contains two distinct galaxy populations. This scenario is qualitatively in agreement with recent phenomenological mass-quenching models and extends them to span more than three quarters of the history of the Universe. 
\end{abstract}


\section{Introduction}

One of the main challenges in observational cosmology is trying to understand the formation and evolution of galaxies based  on their dark and baryonic components, specifically to test theoretical models of galaxy formation at every redshift. 
A large number of surveys have been developed during the last decade to construct multi-wavelength observations of galaxies using diverse ground and space-based facilities. These surveys strive to understand how the galaxy populations observed at early times evolved into those in our local universe. An important feature to be able to model is how the most massive galaxies seen locally have assembled most of their mass and what type of evolution characterized this growth. A critical epoch for galaxy formation was $2\leq z \leq 4$, when star-formation activity in the universe was at its peak, and most of the structures  that we observe in the local universe were not yet present (Shapley 2011). However, a specific evolutionary path from this redshift to the present time is still undetermined (Shapley 2011, McCracken et al. 2010).

Building large statistical samples of galaxies allows us to explore galaxy populations at early epochs and to develop a more complete view of the evolution of structure in our universe. Large infrared (IR) -selected samples are essential for many reasons: 1) spectral features at these important redshifts move out of the optical into the near-IR, 2) long wavelengths are less affected by dust attenuation, and --- especially --- 3) observations at these wavelengths more closely correspond to stellar mass-selected samples since the flux  comes from low-mass, cooler stars that represent most of the mass of the system.

A complete exploration of galaxies at these epochs would require large spectroscopic IR surveys, which are time consuming and realistically unfeasible. Colour-colour selection techniques serve as an efficient way of constructing large samples of galaxies. These selection methods have been developed to distinguish between star-forming and passive galaxies. This distinction can be used to infer information that can be useful for understanding the evolution of the two populations.

Some examples of such colour-colour criteria used to select high-redshift galaxies are the Lyman Break Galaxies (Steidel et al. 1996, 2003), BX and BM galaxies (Steidel et al.\ 2004), Extremely Red Objects (Elston et al. 1998, Thompson et al. 1999, Roche et al. 2002), Distant Red Galaxies (Franx et al. 2003) and \bzk selection (Daddi et al. 2004). The Lyman break and BX/BM techniques were developed to select $z\sim2 - 3$ star forming galaxies, but miss many passively evolving and/or heavily dust-reddened objects. The Extremely Red Objects (EROs) and Distant Red Galaxies (DRG) techniques are designed to select passive high-redshift galaxies based on their red colours. However, spectroscopic follow-up shows that these samples include both passive galaxies and dust-reddened star-forming galaxies (Cimatti et al. 2002, van Dokkum et al. 2004). Of all of these techniques, the \bzk selection technique is the only one able to select and distinguish between star-forming and passive galaxies at $z\sim2$ (Daddi et al. 2004, Grazian et al. 2007). This ability provides a way to trace the evolution of these different populations to high redshift, thereby providing a more complete picture of galaxy formation and evolution.  Consequently, the \bzk\ technique has enjoyed much popularity in recent years (e.g., Kong et al. 2006, Lane et al. 2007, McCracken et al. 2010, Bielby et al. 2012). However, despite this prominence even such a basic property of the \bzk\ population as its luminosity function is still only poorly constrained, and especially so for passive \bzk\ galaxies. 

In this work we adapt the \bzk\ selection criteria to the filter set used in the optical and IR observations of the Canada-France-Hawaii Telescope Legacy Survey (CFHTLS).  We apply our selection criteria to these large fields (0.4--0.9 $deg^{2}$ each) and independent lines of sight in the four Deep Fields of the CTHTLS to construct large, statistically representative samples of star-forming and passive galaxies at \zs2. We then constrain the number counts, rest-frame R-band luminosity functions (LFs), and stellar mass functions (SMFs) of both the star-forming and passive populations. The four separate lines of sight and large effective areas provide low statistical uncertainties while at the same time monitoring the field-to-field fluctuations due to cosmic variance.

Throughout this work we use a flat lambda cosmology ($\Omega_{m}=0.3$, $\Omega_{\Lambda}=0.7$) with $H_{0}$ = 70 km s$^{-1}$ Mpc$^{-1}$. Unless stated otherwise, we use the AB magnitude system (Oke 1974) and the Salpeter (1955) stellar Initial Mass Function (IMF) throughout. 


\section{Data}\label{sec:data}

Our data consist of a combination of two deep wide-field public datasets, one optical and one infrared, taken primarily with the 3.6m Canada-France-Hawaii Telescope (CFHT) on Mauna Kea.  The optical data are the four deep fields of the CFHT Legacy Survey (The CFHTLS T0006 Release, Goranova Y., et al. 2009); the IR data are from the deep imaging in these CFHTLS fields done by the WIRCam Deep Survey (WIRDS; Bielby et al. 2012). Together, these four fields, referred to as D1, D2, D2, and D4, cover a total area of 2.47 $deg^2$ and have significant overlap with regions of the sky that have been extensively studied in other surveys, including COSMOS (D2) and the Groth Strip (D3); see, e.g., Gwyn (2011) and references therein.  Each of the four fields is large and widely separated on the sky  to sample variance (cosmic variance). This can be expected to be a significant contributor to any observed differences between the fields. 

Although we are aware of the existence of other datasets, such as UltraVISTA and VIDEO, which may provide better coverage of some fields or better depths. These datasets were not available at the time this paper was being developed so we have not included them in our analysis.

\subsection{Optical Data: the CFHTLS Deep Fields}

The CFHTLS is a large project carried out by Canada and France that used 50$\%$ of the dark and grey telescope time from mid-2003 to early 2009. For the present work, we used the Deep Survey component of the CFHTLS, which was originally developed to detect 500 type Ia Supernovae and to study the galaxy distribution down to a limiting magnitude r'=28.  

The CFHTLS was carried out using MegaCam which is the wide-field optical imager at MegaPrime, the wide-field facility at CFHT. MegaCam has 36, 2048$\times$4612 pixel CCDs covering a 1~deg $\times$ 1~deg field-of-view with a total of 340 Megapixels with a resolution of 0.186\arcsec\ per pixel. The CFHTLS Deep Survey consists of four widely-separated 1~deg$^2$ fields --- labelled D1, D2, D3, D4 --- each of which was observed in five broad-band filters (u*, g', r', i', z'). Each one of these with a $80\%$ completeness (stellar/compact objects) of 26.22$\pm$0.10, 25.94$\pm$0.10, 25.40$\pm$0.10, 25.10$\pm$0.10,  and 25.59$\pm$0.10 respectively\footnote{Strictly speaking, the CFHTLS i'-band observations used two slightly different i' filters, but since we do not use i'-band data in this work we do not discuss this issue further.}. These filters were designed to match the Sloan Digital Sky Survey (SDSS) filters as closely as possible except for the u* filter, which is designed to take advantage of Mauna Kea's lower UV extinction than that at the Apache Point site of the SDSS. 

We used the CFHTLS T0006 data release which contained two sets of images, one with the 85\% best-seeing images, and the other with the 25\% best-seeing images.
While the 85\% image stacks are deeper than the 25\% stacks for extended sources, our tests showed that the difference was small for compact galaxies such as our high-redshift objects. Consequently we chose to use the 25\% images as these also give us the potential to better investigate morphologies or close companions in future work.

Several entities were involved in the acquisition and processing of CFHTLS data: CFHT was used for data acquisition and calibration while the pre-processing of the images was done by Terapix\footnote{Traitement \'El\'ementaire R\'eduction et Analyse des PIXels, Institut d'Astrophysique de Paris, http://terapix.iap.fr}. Pre-processing involves image quality checking, flat fielding, stacking from dithered images, identification of bad pixels, removal of cosmic rays and saturated pixels, background estimation and subtraction, and astrometric and photometric calibration.

\subsection{Infrared Data: WIRDS} 

The WIRCam Deep Survey (WIRDS) is a large project that obtained deep $J$, $H$ and $K_{s}$ imaging of large subareas of the four CFHTLS Deep Fields. WIRDS was carried out 
between 2006 and 2008, with the bulk of the data being obtained at CFHT using WIRCam (Marmo, 2007), but one image (the $J$-band image of the D2 field) was obtained using the WFCAM instrument on UKIRT.

WIRCam has four detectors in a 2$\times$2 array with 2048$\times$2048 active pixels and covers a 20 arcmin $\times$ 20 arcmin field-of-view with a sampling of 0.3 arsec per pixel. The WIRCam images have been re-sampled to match the 0.186 arcsec/pixel MegaCam pixel scale. 

Pre-processing of the WIRDS data includes a reduction process as follows: bias subtraction and flat-fielding, initial sky subtraction, cross-talk correction, second-pass sky-subtraction, astrometric and photometric calibration and the production of final processed image stacks. For this project we used the T0002 WIRDS data release. For the details of the WIRDS data processing see Bielby et al. (2012). 

\subsection{Object Detection and Photometry}

\begin{table*}
	\centering
		   \caption{Summary of our four fields. Effective areas are given in $deg^{2}$ after removing masked regions. Also shown is the colour excess \ebv\ at the centre of each field and the number of objects found to $K_s=23.0$.  When a range is given, the lower number indicates the number of galaxies that are unambiguously at \zs2, while the higher number includes those objects that cannot be ruled out as low-$z$ interlopers. 		   		   }
		\begin{tabular}{|c|c|c|c|c|c|c|}
		\hline
		Field & Effective Area [$deg^{2}$] & $E(B-V)$ & All objects & gzHK & PE-gzHK & SF-gzHK\\\hline
		D1 & 0.68 & 0.0254  &55,256 & 11,258--12,281 & 1,382 & 9,972--11,004\\
		D2 & 0.89 & 0.0162 & 87,206 & 12,238--15,222 & 1,739 & 10,880--13,835\\
		D3 & 0.45 & 0.0072  & 37,380 & 7,046--7,668 & 841 & 6,223--6,845 \\
		D4 & 0.45 & 0.0275  & 38,461 & 7,312--8,168 & 1,013 & 6,361--7,226 \\
		  \hline
		\end{tabular}
	\label{tab:catalogs2}
\end{table*}

We used SExtractor version 2.8.6 (Source Extractor, Bertin and Arnouts 1996) for object detection and photometry.  Object detection was done in the unsmoothed $K_s$ images.  SExtractor parameters were tuned to detect the vast majority of objects identifiable by eye without producing significant numbers of spurious detections; in practice this meant that we required that an object should have a minimum of 5 contiguous pixels above the 1.2$\sigma$ sky level in the WIRDS $K_s$ image.  Kron-like apertures, in which the first image moment is used to determine the flux of the galaxy from a circular/elliptical aperture, (Graham et al. 2005)  were used to determine the total magnitude of a galaxy.

To ensure accurate colours, object colours were measured on PSF-matched images through matched apertures.  The PSF matching involved local PSF determination within the images, followed by smoothing to match the local PSF in the worst-seeing image in each field. It takes into account seeing variations across the large, multi-chip mosaics used here. For full details of the PSF-matching see the paper by Sato \& Sawicki (2013).  Fluxes within 1.86\arcsec\ (10 pixel) -diameter apertures were then determined at the positions of the $K_s$-band detections using SExtractor's dual-image mode, and these fluxes were then combined to produce object colours. 

Corrections for extinction due to dust in our Galaxy were made based on the Schlegel, Finkbeiner, \& Davis (1998) dust maps, using the $E(B-V)$ value at the centre of each field (see Table~\ref{tab:catalogs2}) to correct the magnitudes and colours of all the objects in that field. Because extinction is small in the CFHTLS Deep Fields, dust corrections are likewise small and using the central field values is a reasonable approximation. 

Next, we masked out suspect areas of the images from our final science catalog. These areas consist of regions likely to yield spurious detections or suspect photometry and include areas near stellar diffraction spikes, bleeding columns, residual satellite trails, or reflective halos from bright stars caused by reflections from the CCD and instrument optics. Unexposed areas (based on weight images supplied by Terapix), are also excluded.  Additionally, using SExtractor's internal flags, we are able to identify saturated and/or truncated objects in our sample. Rejecting these objects flagged by SExtractor does not represent a considerable loss of data (approximately only 1.6$\%$ of our sample), but ensures the removal of objects whose colours may be inaccurate.


Table \ref{tab:catalogs2}  gives a summary of the four Deep fields, the resulting effective area after the masking of suspect regions, the colour excess at the field centre, and the number of objects found in each field to $K_s$=23.0.


\section{The \gzhk Selection and Classification of \zs2 Galaxies}
\label{sec:gzhk}
 
Daddi et al. (2004) developed the \bzk selection technique to select \zs2 galaxies and classify them into star-forming and passively evolving objects.  The CFHTLS+WIRDS g', z', and \Ks filters differ from the Daddi et al. filters and so it is necessary to adjust our approach to account for that. While some studies transform the colours of their observed objects to bring them onto the \bzk system (e.g., Kong et al. 2006, Blanc et al. 2008, Hartley et al. 2008, McCracken et al. 2010), we use a different approach in that we work in the native g'z'\Ks\ filter set of our data and instead adapt the colour selection cuts to match this filter set, as described in \S~\ref{sec:bzk-to-gzk}.

One strength of the \bzk technique is its ability to distinguish between \zs2 star-forming and passive galaxies.  However, because a typical passively-evolving galaxy at \zs2 is expected to have very red colours (typically  $B-K > 5$), this approach requires extremely deep observations at blue wavelenghts. Since we wish to probe as deep into the \zs2 luminosity functions as our \Ks-band selection allows (\Ks $\lesssim$ 23--24), and since our optical data are not particularly deep in comparison, in \S~\ref{sec:zhk} we develop a modification to the \gzk technique that, by incorporating $H$-band photometry, allows us to reliably discriminate between star-forming and passive galaxies without the need for ultra-deep blue data. We thus follow a two-step process: \gzk selection to select \zs2 galaxies without knowledge of their spectral type (\S~\ref{sec:z2-with-gzK}) followed by \zhk-based classification into star-forming and passive populations (\S~\ref{sec:zhk}).

A similar analysis was performed to ensure that our z-band was deep enough to avoid low-redshift interlopers or mistakenly classify star-forming galaxies as passive. We found that contamination of the blue population by $z$-band non-detections is negligible, while that of the red population is only about $3\%$

To model galaxy colours, following the colour-colour criteria of Daddi et al. (2004), we use rest-frame model spectra  of constant-SFR and single-burst stellar populations from the GALAXEV library (Bruzual and Charlot 2003). Using solar metallicity models and the Salpeter (1955) stellar initial mass function, we apply attenuation by interstellar dust (Calzetti et al.  2000), redshifting to the observed frame, and integration through the filter transmission curves using the SEDfit package (Sawicki 2012a).


\subsection{Galaxy Models in the \bzk  and \gzk  Diagrams}
\label{sec:bzk-to-gzk}

\begin{figure*}
\centering
     \includegraphics[width=15.2cm]{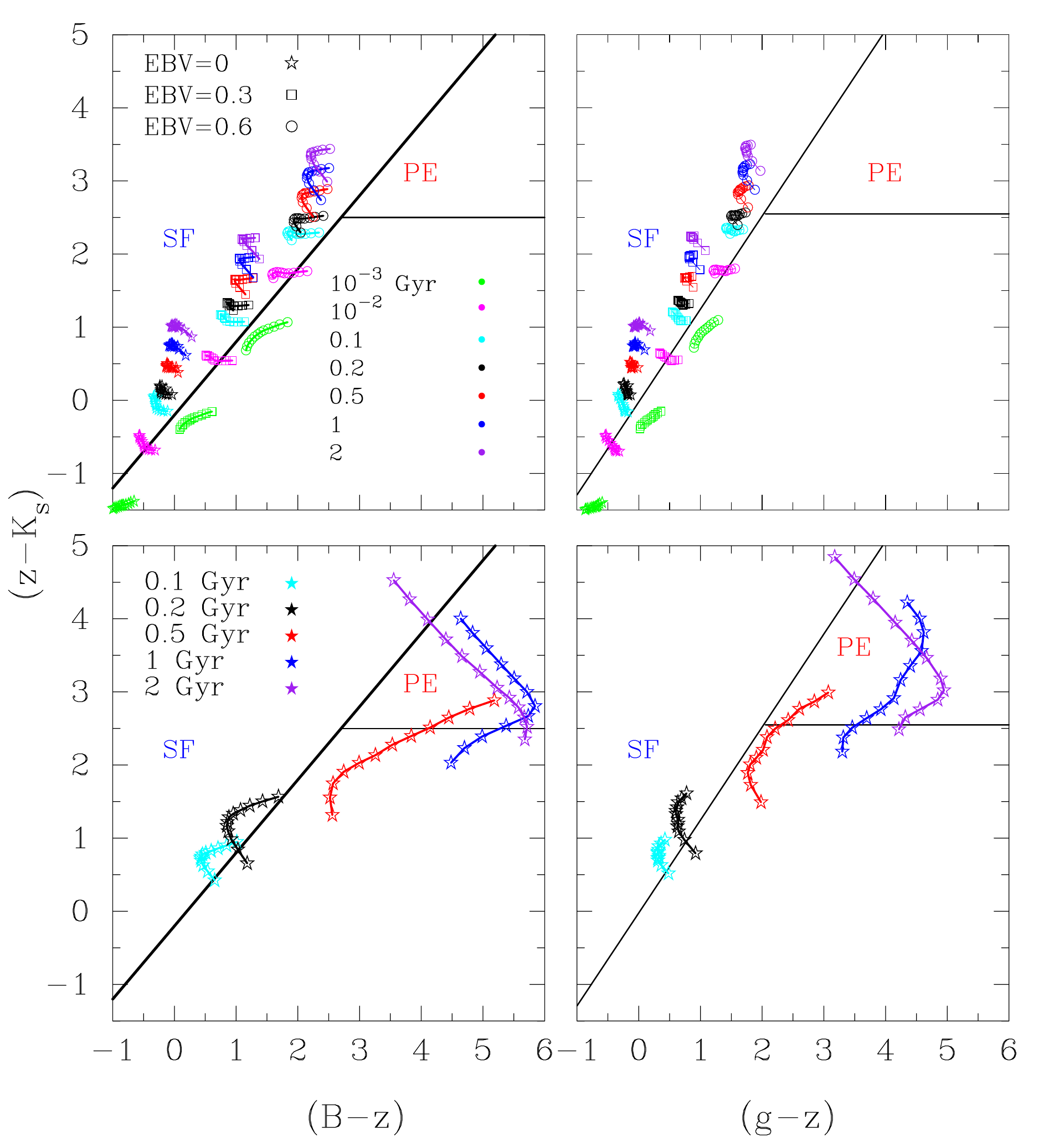} 
   \caption{Bruzual and Charlot (2003) stellar population synthesis models for \bzk and \gzk colour-colour plots with 1.4$\leq$z$\leq$2.5. Top-left panel :  These models are intended to reproduce \bzk colours of star-forming galaxies. Ages in Gyr are shown with different colours while different E(B-V) are represented with different symbols. Top-right panel: The same constant star-formation (CSF)  models shown previously in the \bzk plot, are simulated the CFHT \gzk filters. Bottom-left panel: Instantaneous Burst (SSP) Models. These models are intended to reproduce \bzk colours of passive galaxies with different ages and E(B-V)=0. Bottom-right panel: Shows the same SSP models shown in the bottom-left panel but in a \gzk colour-colour plot. The solid black lines on the left-side plots are the same as given by Daddi et al. (2004). Solid black lines in the right-hand side plots are shifted from the original criteria to select the same \bzk models.}
   \label{fig:multigzk-bzk}
  
\end{figure*}

Daddi et al.\ (2004)  developed a highly popular technique that allows both the selection and classification of \zs2 galaxies using a simple \bzk\ colour-colour diagram. Reproducing the \bzk\ selection diagram of Daddi et al. (2004), in the top-left panel of Figure~\ref{fig:multigzk-bzk} we plot star-forming Bruzual and Charlot (2003) constant-SFR (CSF) models with ages between $10^{-3}$ and 2 Gyr, solar metallicity, Salpeter stellar initial mass function (IMF), $0 \leq E(B-V) \leq 0.6$, and 1.4 $\leq z  \leq $ 2.5. As shown in the bottom-left panel of Figure~\ref{fig:multigzk-bzk}, passive galaxies are represented using instantaneous burst models (SSP) with older stellar populations (ages between 0.1 and 2 Gyrs), and no dust. In the \bzk technique high-redshift objects are uniquely located in regions of the \bzk diagram: objects that lie to the left of the diagonal line are considered to be star-forming \zs2 galaxies, while those to the right of the diagonal line but above the horizontal line are considered \zs2 passively-evolving galaxies.  Spectroscopy, along with morphological studies, support the validity of this approach (Daddi et al. 2004, 2005,  Ravindranath et al. 2008, Mancini et al. 2009, Onodera et al. 2010).

The location of galaxies in the color-color diagram (Fig.~\ref{fig:multigzk-bzk}) may be affected by the choice of IMF, dust models, or other model parameters. Such effects could affect the selection of high-z galaxies as well as their classification into passive and star-forming populations, and we plan to investigate the scope of these effects in future work; in the present paper we simply use the models as described above since they provide consistency with the basic Daddi et al. (2004) technique and the many papers the follow it.

To modify the \bzk technique for use with the CFHTLS+WIRDS \gzk filters we considered the positions of the Daddi et al. (2004) models (left panels of Fig.~\ref{fig:multigzk-bzk}) in the \gzk colour-colour diagram (right panels of Fig.~\ref{fig:multigzk-bzk}).  While the \gzk and \bzk filter systems are different, they are similar enough that the main qualitative features of the colour-colour diagram are preserved, as can be seen by comparing the left and right panels of Fig.~\ref{fig:multigzk-bzk}.  We conclude that even though the filters are somewhat different, we can adjust the original Daddi et al.\ (2004) \bzk selection to develop a \gzk colour-colour selection.

In analogy with the \bzk technique, we define lines in \gzk colour-colour space that delineate regions inhabited by different types of galaxies. These colour-colour cuts are defined by identifying the models that in the \bzk space lie close to the Daddi et al. (2004) \bzk criteria, locating the corresponding models in the \gzk colour-colour space, and then designing simple colour cuts in \gzk space in analogy to those in the \bzk diagram.
Our adopted \gzk colour cuts are shown with black lines in the right-hand panels of Fig.~\ref{fig:multigzk-bzk} and are described as follows. Star-forming \gzk galaxies can be selected using
\begin{equation}
\label{eq:sfgzk}
(z-K_{s})-1.27(g-z)\geq -0.022, 
\end{equation}
and passively evolving ones via
\begin{equation}
\label{eq:pegzk}
(z-K_{s})-1.27(g-z)< -0.022 \hspace{0.1cm} \cap \hspace{0.1cm} (z-K_{s})>2.55.
\end{equation}
Finally, in analogy with the \bzk technique, by visual inspection of the data we identify the stellar locus and distinguish between stars and galaxies using
\begin{equation}
\label{eq:stgzk}
(z-K_{s})-0.45(g-z)\leq-0.57.
\end{equation}
It is important to note that our high-redshift selection criteria (Equations ~\ref{eq:sfgzk} and \ref{eq:pegzk}) are designed in close reference to the \bzk criteria of Daddi et al. (2004) via the comparison of model locations in the two colour-colour spaces. Consequently, the galaxy populations selected by our new \gzk criteria can be expected to closely match those selected using the traditional \bzk technique. 


\subsection{Selection of \zs2 Galaxies in the CFHTLS Deep Fields}
\label{sec:z2-with-gzK}

In principle, our \gzk selection criteria (Equations~\ref{eq:sfgzk} and \ref{eq:pegzk}) allow us to select and classify high redshift galaxies in a way that closely resembles the popular \bzk technique of Daddi et al. (2004) but is directly applicable to the \gzk CFHTLS+WIRDS filter set. However, as can be seen in Figure~\ref{fig:multigzk-bzk} high-redshift galaxies are expected to have very red colours, so the use of \gzk (or \bzk) selection alone requires very deep g' (or $B$) data. With $K_{s, lim} \sim$ 23--24, we would require optical data reaching g'$\sim$28--29 in order to discriminate between passive and star-forming \zs2 galaxies. The current CFHTLS data, as deep as they are for such a wide survey, are not deep enough.  

\begin{figure}
   \centering
   \includegraphics[width=7.2cm]{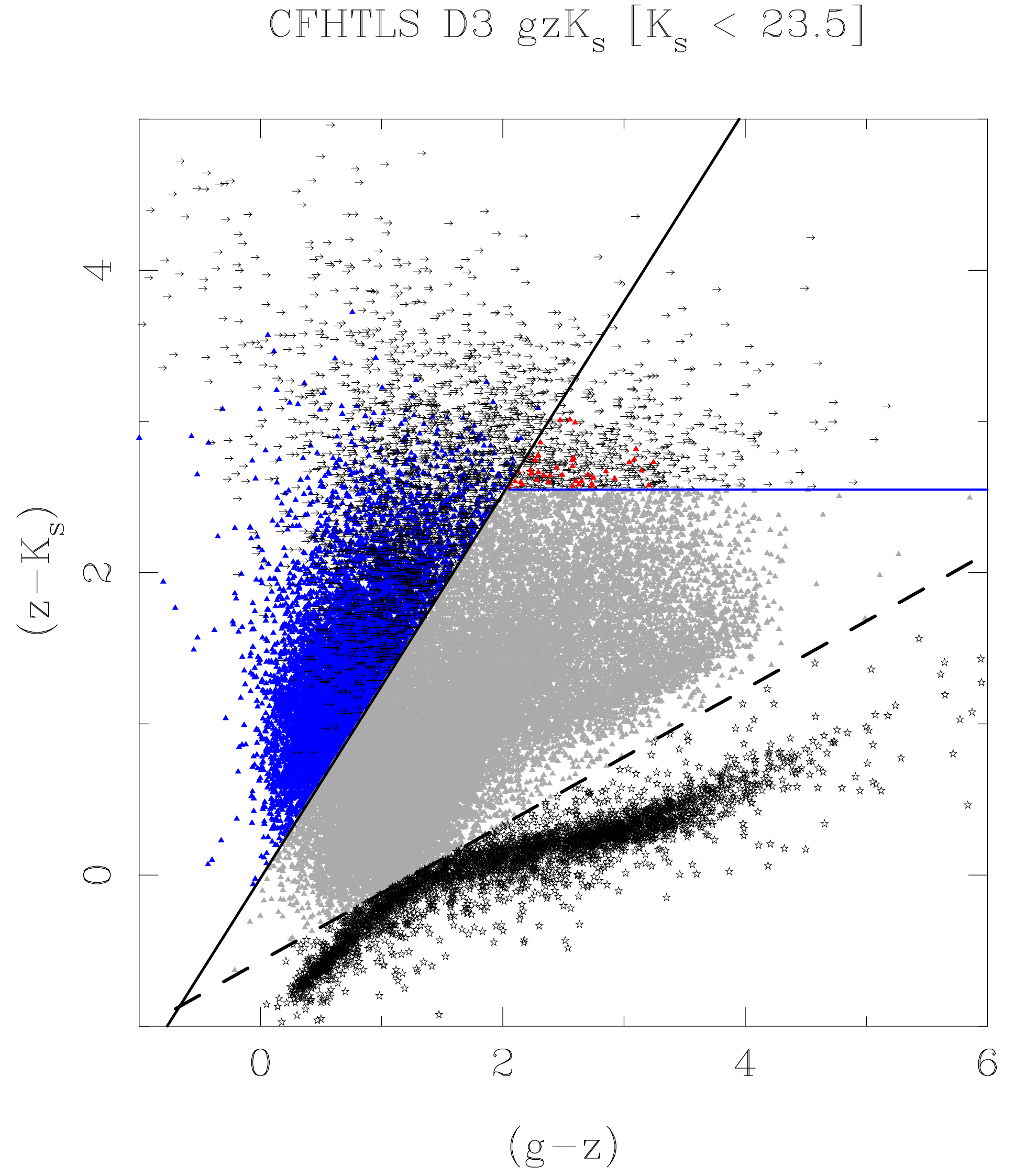} 
   \caption{Objects in field D3 down to $K_s$ = 23.5. Objects detected in g and classified as star-forming high-redshift galaxies are represented by blue triangles, while those detected in g and classified as passive galaxies are shown with red triangles. Arrows represent 1$\sigma$ limits for objects that have not been detected in g.  Stars reside in the lower-right of the diagram and are shown with star symbols. The solid lines represent the colour cuts defined by Equations~\ref{eq:sfgzk} and \ref{eq:pegzk}, and the dashed line shows the stellar cut of Equation~\ref{eq:stgzk}.}
   \label{fig:gzk}
\end{figure}

This limitation is illustrated in Fig.~\ref{fig:gzk}.  Although our data allow us to identify all objects with z$-K_s>$ 2.55 as high-redshift (\zs2), the majority of these galaxies are undetected in g' and so we cannot tell whether they are passive or star-forming \zs2 objects. We also face a lesser, but related problem in that a fraction of the galaxies with z$-K_s<$ 2.55 is undetected in g', making it unclear whether they are \zs2 galaxies or low-redshift objects. One solution to these problems would be to obtain much deeper g' data than those already in hand, but this approach would be very costly in telescope time. Instead we take a different approach, as follows. 

While unable to usefully distinguish between star-forming and passive galaxies, the present data are deep enough to select \zs2 galaxies (albeit with redshift ambiguity for the relatively small number of g'-undetected objects below $(z-K_s)=2.55$).  In other words, the union of the regions defined by Equations~\ref{eq:sfgzk} and \ref{eq:pegzk}, 
\begin{equation}
\label{eq:allgzk}
(z-K_{s})-1.27(g-z)< -0.022 \hspace{0.1cm} \cup \hspace{0.1cm} (z-K_{s})>2.55,
\end{equation}
selects all \zs2 galaxies so long as they are detected in g'.  

We thus adopt Equation~\ref{eq:allgzk} as our way to select \emph{all} \zs2 galaxies. By either including or excluding the g' non-detections in our accounting, we then construct two conservative samples: one sample includes all objects that could be star-forming \zs2 galaxies (but with some lower-z contamination);  the other consists only of objects that are sure to be at \zs2 star-forming galaxies, but is not complete in the sense that it lacks those \zs2 star-formers that are undetected in g'.  The number of galaxies in these samples is listed in Table~\ref{tab:catalogs2} under the heading ``\gzhk".

The above approach does not give us information on the evolutionary states of our \zs2 galaxies. Consequently, to determine whether these galaxies are star-forming or passive we develop a new technique, described in \S~\ref{sec:zhk}.

 \subsection{Classification into passive and star-forming populations using \zhk colours}
 \label{sec:zhk}

In \S~\ref{sec:z2-with-gzK} we defined samples of \zs2 galaxies in the CFHTLS Deep Fields.  All \zs2 galaxies below \zk\ $=2.55$ are by definition star-forming, but we were unable to determine the evolutionary state of the \zs2 galaxies above \zk\ $=2.55$ using the \gzk technique alone because we lack sufficiently deep g' data. To address the passive/star-forming ambiguity above $z-K_{s}=2.55$, we now develop and apply a follow-up selection technique based on (z$-H$) versus ($H-K_{s}$) colours.

\begin{figure}
   \centering
   \includegraphics[width=7.3cm]{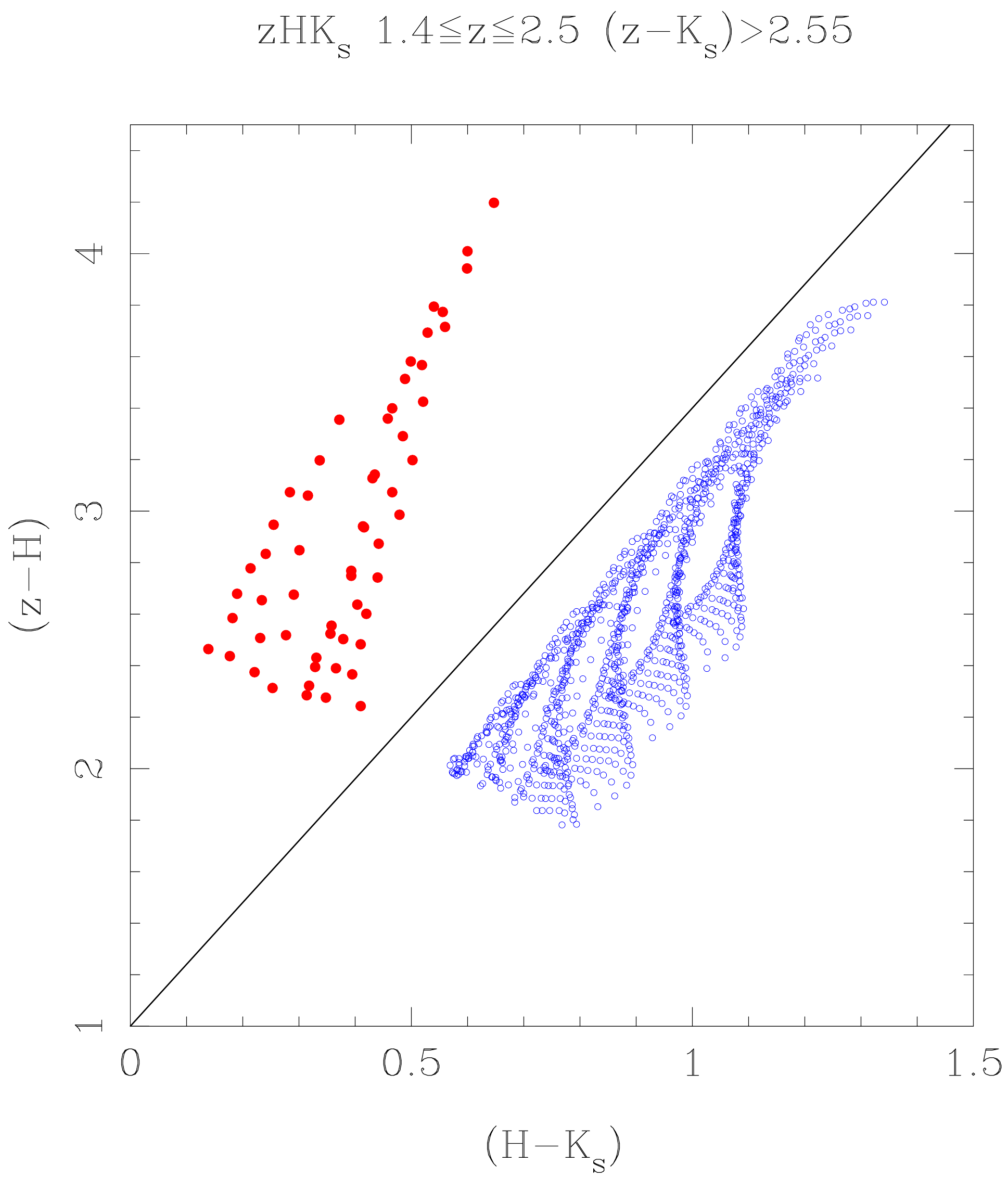} 
   \caption{Model (z-H) versus (H-$K_{s}$) colours for those objects with $(z-K_s)>2.55$ in the \gzk diagram (see Figure \ref{fig:multigzk-bzk}).  Blue points represent star-forming galaxies with $log_{10}(Age)$ between 7.28 and 9.36 ($0.02$ and 2 Gyr) and E(B-V)$\geq$0. Red points represent passive galaxies with $log_{10} (Age)$ between 8.63 and 9.36 (0.4 and 2 Gyr) and E(B-V)=0. The solid line separates the star-forming and passive regions described by Equations~\ref{eq:sfzjhk} and \ref{eq:pezjhk}.      }
   \label{fig:zhkfit}
\end{figure}

\begin{figure}
   \centering
   \includegraphics[width=7.3cm]{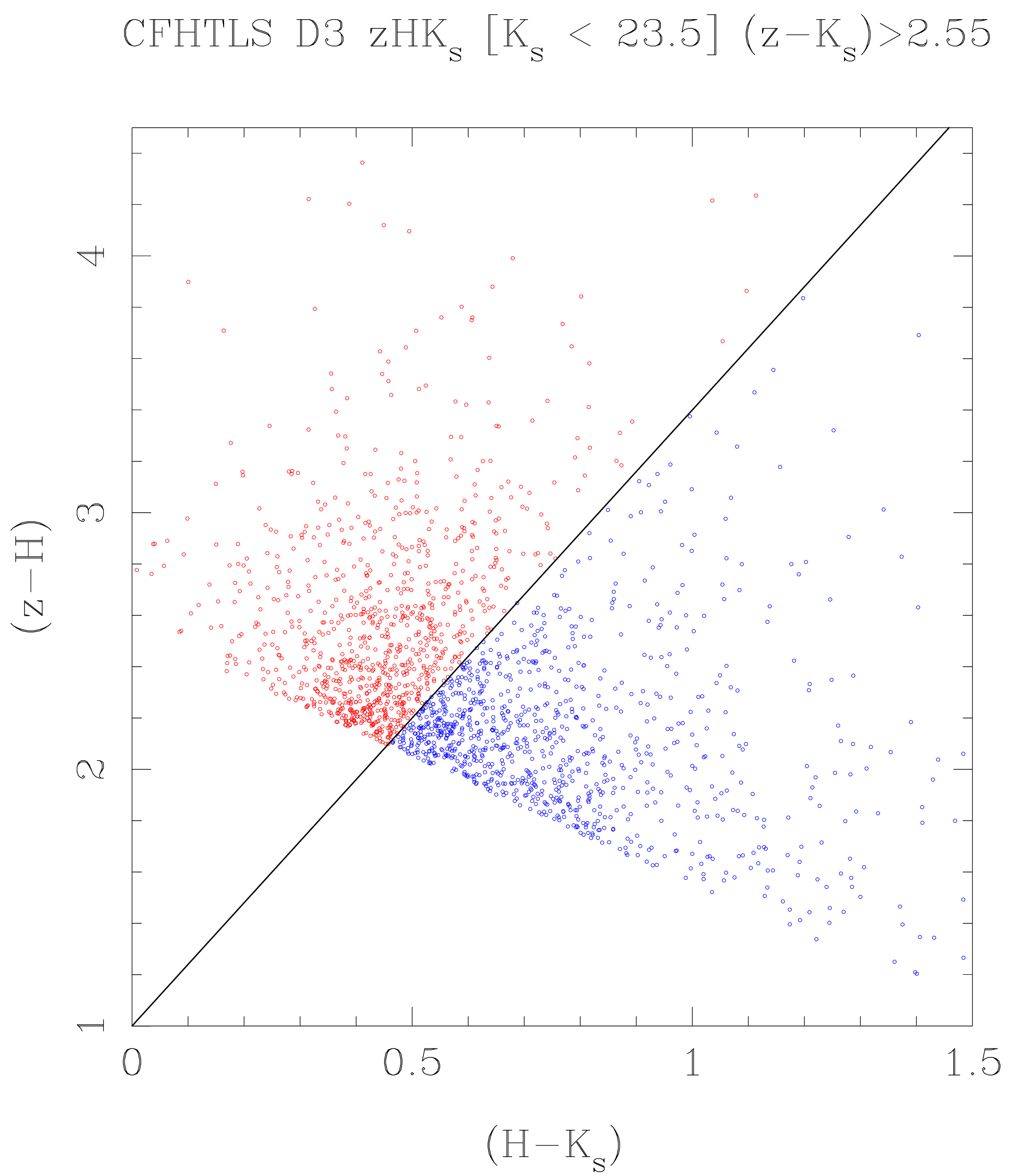} 
   \caption{(z-H) versus (H-$K_{s}$) colours for those objects with $(z-K_s)>2.55$ in CFHTLS Field D3 to $K_{s}<23.5$.  Blue points represent star-forming galaxies while red points represent passive galaxies. The solid line separates the star-forming and passive regions described by Equations~\ref{eq:sfzjhk} and \ref{eq:pezjhk}. The diagonal cut-off in the lower part of the diagram reflects the $z-K_s > 2.55$ colour cut.     }
   \label{fig:zhk}
\end{figure}

The red colour of $(z'-K_{s})>2.55$ galaxies is caused either by dust (in the case of star-forming objects) or by the presence of the 4000\AA/Balmer break complex (for sufficiently old passive systems).  Dust attenuation of star-forming spectra produces a continuous, tilted spectral slope, while, in contrast,  the  4000\AA/Balmer break complex gives a sharp spectral discontinuity. Consequently, it should be possible to distinguish between the two phenomena, and therefore between dusty star-formers and passive systems, by adding a flux measurement between the z' and $K_s$ bandpasses. Specifically, at $z\sim2$ the 4000\AA/Balmer break complex is redshifted to $\sim$11,000--12,000\AA\ and so will fall between the z' and $H$ filters.  Therefore, we should be able to identify this feature using the $(z-H)$ versus $(H-K_{s}$) colour-colour diagram (Figure~\ref{fig:zhkfit}).

The passive/star-forming ambiguity exists for  $(z'-K_{s})>2.55$ objects only, and so in Figure \ref{fig:zhkfit} we plot only those models from the \gzk diagrams (Figure~\ref{fig:multigzk-bzk}) that are above $(z-K_{s})=2.55$.  As expected, the star-forming and passive models lie in different regions of Figure~\ref{fig:zhkfit} and thus provide us with the means to distinguish the two populations. 

Note that for a star-forming galaxy, as age, dust, and redshift are increased, the models shift parallel to the diagonal line shown in the Figure~\ref{fig:zhkfit}.  This linear shift occurs because increasing dust will affect both colours by roughly the same amount and, moreover, the fact that the star-forming spectrum is not expected to have a break also ensures smooth evolution with redshift.  In essence, star-forming galaxy spectra, even in the presence of dust, are essentially power laws at these wevelengths, ensuring their position in the lower right of Figure ~\ref{fig:multigzk-bzk}.  

In contrast, for passively evolving models the presence of a spectral break results in a much stronger change between the (z-H) and (H-$K_{s}$) colours. As expected, models that represent passively evolving galaxies lie on a different locus in Figure~\ref{fig:zhkfit}, where they can be clearly separated from the constant star formation models.  We adopt the solid black line in Figure \ref{fig:zhkfit} as the division between models of star-forming and passive \gzk-selected galaxies that have $(z-K_{s})>2.55$.  
 
We then use the following criteria to select and classify \zs2 galaxies in our sample.  First, as described in \S~\ref{sec:z2-with-gzK}, we use Equation~\ref{eq:allgzk} to select \zs2 galaxies ragardless of their spectral type. Below $(z-K_{s})=2.55$ all objects selected using Equation~\ref{eq:allgzk} are classified as star-forming \zs2 galaxies.  Above  $(z-K_{s})=2.55$, all galaxies selected using Equation~\ref{eq:allgzk} are also deemed to be at \zs2, and are the classified as star-forming if 
\begin{equation}
\label{eq:sfzjhk}
(z-H)\leq2.4(H-K_{s})+1, 
\end{equation}
and as passively evolving if 
\begin{equation}
\label{eq:pezjhk}
(z-H)>2.4(H-K_{s})+1.
\end{equation}
We call the star-forming objects thus selected ``SF-gzHK" galaxies, and the passively evolving ones ``PE-gzHK" galaxies. The models used in designing the above selection criteria are the same as those in the ``classic" \bzk\ colour-colour cuts, and so we stress that the populations selected by our criteria are essentially the same as those selected using the ``classic" \bzk\ technique. 

Application of the above criteria to our photometric catalogs of \S~\ref{sec:data} yields samples of \zs2 star-forming and passively-evolving galaxies in the CFHTLS Deep Fields.  Their bulk properties are summarized in Table~\ref{tab:catalogs2}.


\section{$K_s$-Band Galaxy Number Counts}
\label{sec:counts}

In this section we determine galaxy number counts for different populations of galaxies: star-forming and passive galaxies at \zs2 (as defined by Eq.~\ref{eq:sfzjhk} and \ref{eq:pezjhk}), as well as for all galaxies irrespective of redshift (Eq~\ref{eq:stgzk}). To do this, we count galaxies in 0.5 mag-wide total $K_{s}$ magnitude bins and --- because some faint galaxies can be undetected above night-sky fluctuations --- apply incompleteness corrections determined from simulations.  

Our incompleteness corrections use artificial objects added at random locations into the science images with empirically motivated morphological parameters. Star-forming galaxies are assumed to be disk-like objects with effective radii between $1\leq r_{e} \leq 3$ kpc (Yuma et al. 2011) while passive galaxies present a more compact morphology and effective radii between $3\leq r_{e} \leq 6$ kpc  (Mancini et al. 2010).  The total (i.e., irrespective of redshift) galaxy number counts are corrected using morphologies that mimic the observed CFHTLS galaxy population. Once artificial galaxies are added into an image, object-finding with SExtractor is used to determine the recovery rate as a function of apparent magnitude. 

Incompleteness corrections are typically of the order of $\sim$1.02 over $K_{s}=17-22$ (AB) mag but become larger at fainter levels.  We limit our analysis in magnitude bins where incompleteness corrections are smaller than a factor of two. Table~\ref{tab:completeness} lists our adopted completeness limits. 

\begin{table}
	\centering
		   \caption{Adopted $K_{s}$ band completeness limits.  Listed are bin centres of the faintest 0.5-mag-wide bins that are included in our analysis.}
		\begin{tabular}{|c|c|c|c|c|}
		\hline
		\textbf{Population}&\textbf{D1}&\textbf{D2}&\textbf{D3}&\textbf{D4}\\\hline
		All galaxies&24.0&23.5&24.0&24.0\\\hline
		Star-Forming galaxies&24.0&23.5&24.0&24.0\\\hline
		Passively Evolving galaxies&24.0&23.0&24.0&23.5\\\hline
		\end{tabular}
	\label{tab:completeness}
\end{table}

\subsection{All galaxies, irrespective of redshift}\label{sec:all_numbercounts}

\begin{figure}
   \centering
   \includegraphics[width=7.8cm]{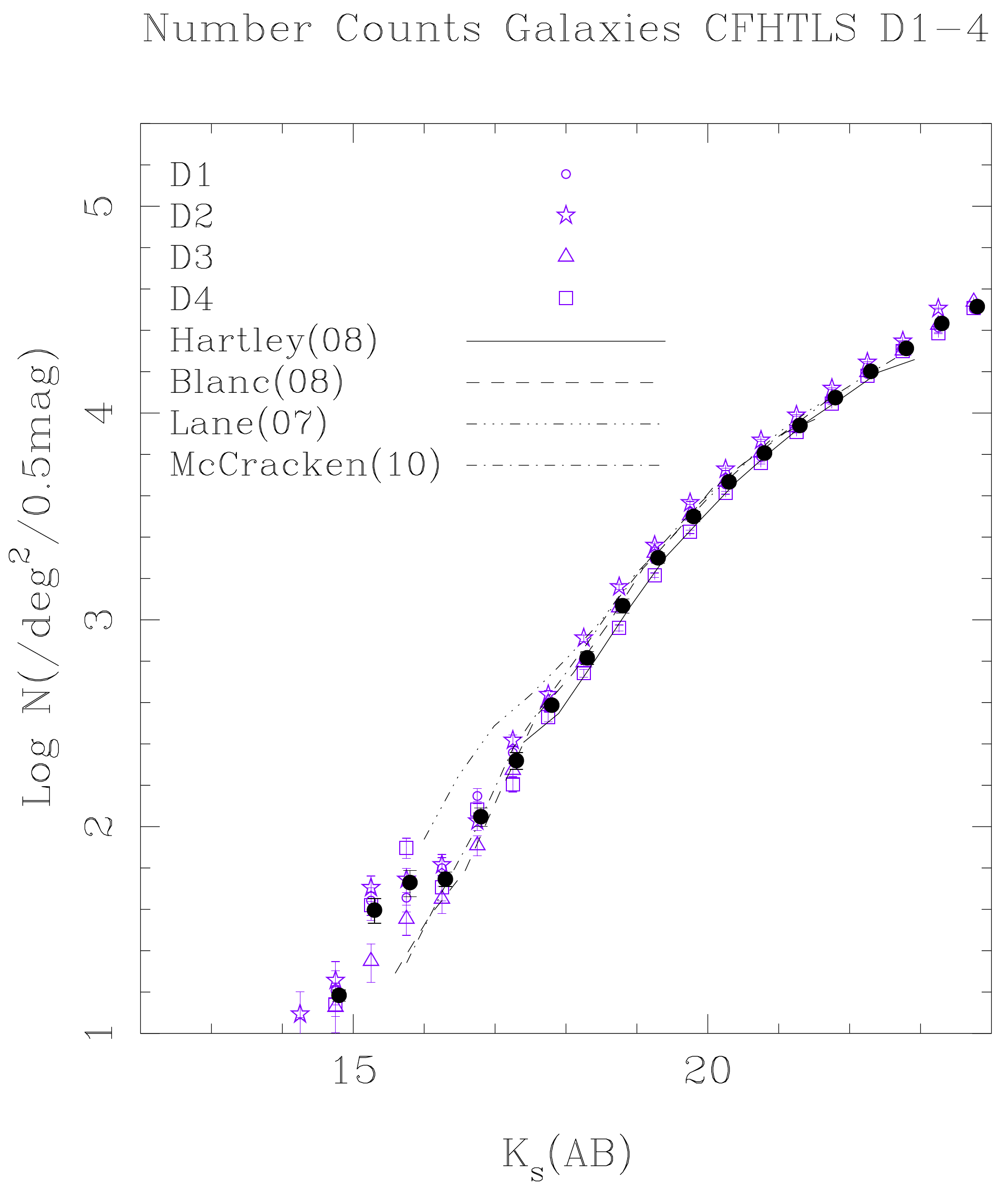} 
   \caption{Number counts for $K_s$-selected galaxies (irrespective of redshift). Our results are in good agreement with other authors and their results of $K_{s}$-selected galaxies.}
   \label{fig:ncall}
\end{figure}

Galaxy number counts for all the galaxies in each field (selected using Equation \ref{eq:stgzk}) are shown in Fig.~\ref{fig:ncall}.  Error bars for each field in this figure show $\sqrt{N}$ uncertainties. Filled black points represent a straight average between the four fields and their error bars represent their sum in quadrature of the individual fields' errorbars.

We compare our results with those of other studies, including Lane et al. (2007), Hartley et al. (2008), Blanc et al.\ (2008), and McCracken et al. (2010)  (We note  that McCracken et al.\ 2010 use the same NIR data as we do, but their survey is limited to the D2/COSMOS field).  Key information on these surveys can be found in Table \ref{tab:authors}.  We do not plot the results of earlier surveys (e.g., Kong et al. 2006, Hayashi et al.\ 2007) because they were limited in either depth or area and their results have been superseded by more recent work.  As can be seen in Fig.~\ref{fig:ncall}, our results at faint magnitudes seem to be in good agreement with the results of the other studies. The small variations between our four fields, as well as between the various surveys, are most likely due to cosmic variance (i.e. variations in galaxy number density due to small-scale inhomogeneities in the universe).

At the bright end, ($K_{s}<18$) there is significant scatter between surveys. Additionally, our counts exhibit a bump around $K_{s}=15$, which could be attributed to a number of reasons: One could be merely due to how SExtractor detects and deals with bright objects. It should also be taken into account that at this bright end, our number counts are relatively low, giving large error bars. Results from Lane et al. (2007) also seem to exhibit a bump at bright magnitudes, while those of McCracken et al. (2010) do not, and the rest of the authors have not presented results at this bright end. Nevertheless, the variations at the bright end are not of great concern here since we are mainly interested in faint objects.

\begin{table*}
	\centering
	\caption{$K_{s}$ Magnitude limits and effective areas for the different authors presented to compare our results. Areas for these different authors are smaller than listed due to masking of bad regions. However, the area presented for this work, CFHTLS+WIRDS is in fact the effective area of the survey after masking of bad regions.}
		\begin{tabular}{|l|c|c|c|}
		\hline
		\bf{Field}&\bf{Mag lim $K_{s}$}(AB)&\bf{Area $deg^{2}$}&\bf{N. of fields}\\\hline
		Lane et al. (2007)&22.50&0.55&1\\
		Hartley et al. (2008)&23.00&0.63&1\\
		Blanc et al. (2008)&21.85& 0.71&2\\
		McCracken et al. (2010)&23.00&1.89&1\\
		This work (CFHTLS+WIRDS)&23.00&2.47&4\\
		\hline
		\end{tabular}
	   	\label{tab:authors}
\end{table*}

\subsection{Star-forming \zs2 galaxies}\label{sec:SF_numbercounts}

\begin{figure}
  \centering
   \includegraphics[width=7.8cm]{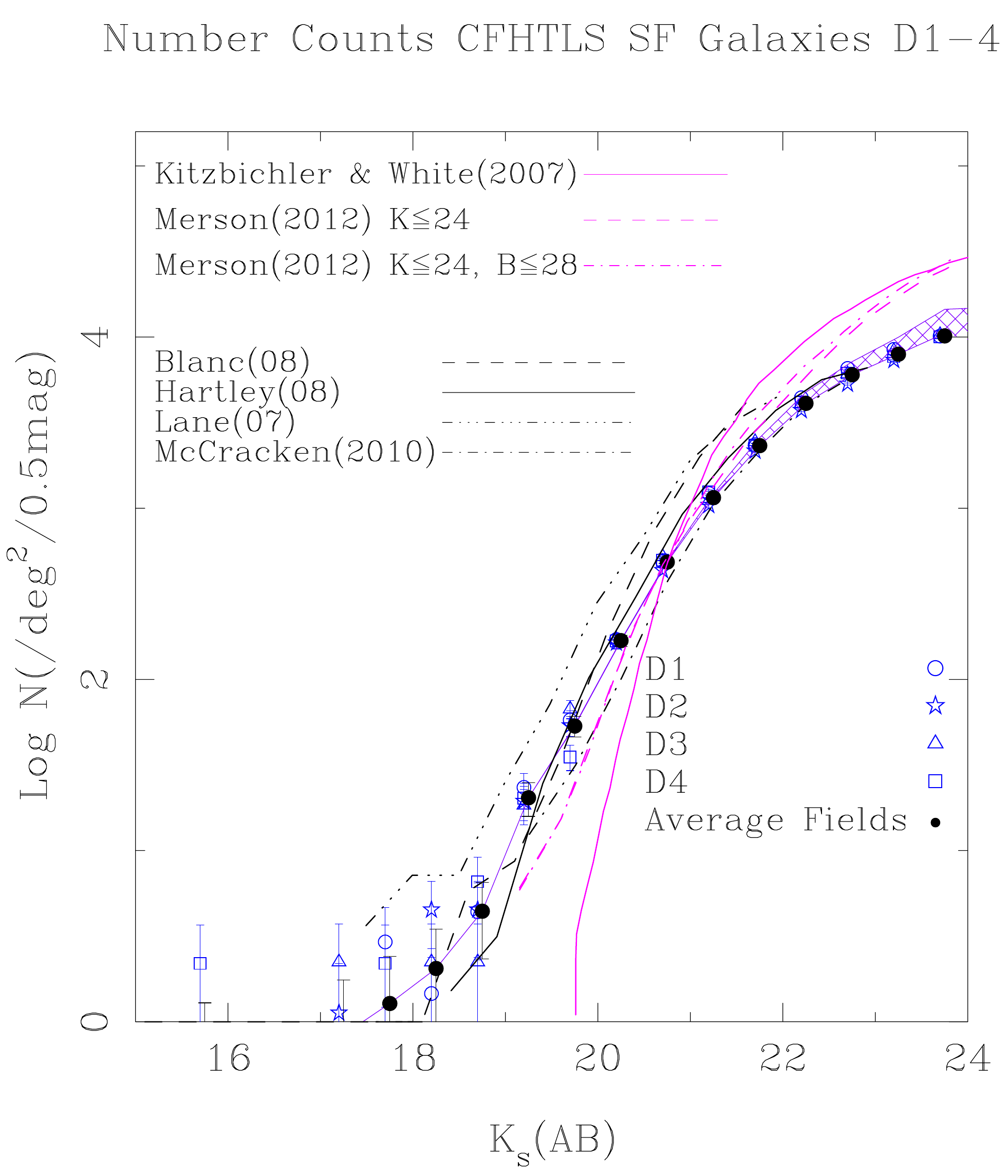} 
   \caption{Galaxy number counts for star-forming galaxies.  Open blue symbols represent our four Deep fields D1-4, with error bars determined from Gaussian statistics. Also shown with filled black circles are galaxy number counts for the average of our four fields, with error bars calculated from the individual fields' error bars combined in quadrature. Both open blue symbols and black filled circles were shifted horizontally from the midpoint of the bin for clarity. As explained in the text, the hatched purple region represents an upper and lower limit between our $g$-detected and $g$-undetected galaxies and thus marks the allowable range of SF galaxy number counts.}
   \label{fig:ncsf}
\end{figure}

Figure \ref{fig:ncsf} shows galaxy number counts for \zs2 star-forming galaxies.   Since many of our SF-\gzhk\ candidates are undetected in $g$-band (\S~\ref{sec:gzhk}), we need to deal with potential low-redshift interlopers amongst the $z-K_s<2.55$ objects (the $z-K_s>2.55$ objects are all regarded as at high redshift and their SF/PE nature is resolved using $zHK_s$ colours alone). To address this interloper issue, Fig.~\ref{fig:ncsf} shows two extreme number count cases. The points show the number counts of  $z-K_s<2.55$ galaxies that are detected in $g$-band and hence known to fulfill the high-$z$ colour selection criteria (in this case we are rejecting from the sample any galaxy that was not detected in g).  As such, the points represent a lower limit on the true number counts of star-forming galaxies.  In particular, the black filled points represent the average of the lower limits set by our four deep fields. In contrast, the purple hatched region  represents the full range of allowable number counts as it accounts for the $g$-band undetected objects, many of which could be low-redshift interlopers. The upper and lower limits of the cross-hatched regions thus represent the allowable range of SF galaxies. Note that this allowed range is quite narrow and thus we can claim to have constrained the number counts of SF galaxies quite well. 

We compare our star-forming galaxy number counts with the counts presented by Blanc et al.\ (2008), Hartley et al.\ (2008), Lane et al.\ (2007), and McCracken et al.\ (2010).  We do not show the $sBzK$ counts of Bielby et al.\ (2012), which are based on the same CFHT data as our work, since their use of traditional three-filter $BzK$ selection in the presence of the relatively shallow CFHTLS $g$-band data can result in misclassifications of PE/SF galaxy spectral types at faint magnitudes.  Our results for star-forming galaxies seem to be in good agreement with most authors except with Lane et al. (2007). Lane et al. (2007) attributed the differences between their number counts of star-forming galaxies and other authors such as Kong et al. (2006)  to cosmic variance. However, cosmic variance between our four CFHT fields, each of which is similar in size to the field of Lane et al.\ (2007), is relatively small: it thus seems unlikely that the very high number counts of Lane et al.\ (2007) should be due to cosmic variance. Blanc et al. (2008) argue that  the Lane at al. (2007) colour selection does not reproduce the standard \bzk\ selection well and our cosmic variance analysis supports that conclusion.

\begin{figure}
   \centering
   \includegraphics[width=7.8cm]{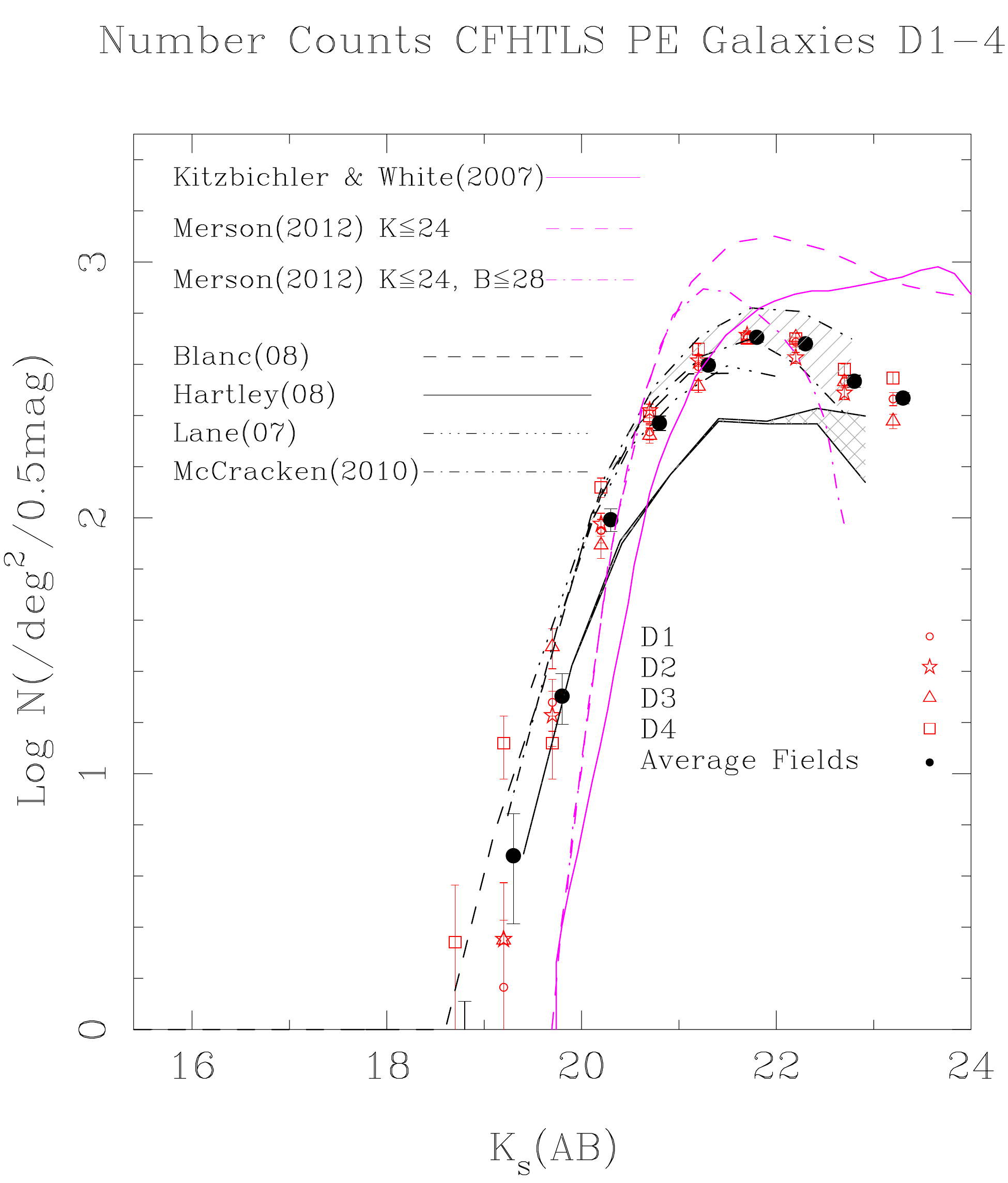} 
   \caption{Galaxy number counts for $z\sim2$ passive galaxies selected using the \gzhk selection criteria.  Open red symbols represent our four Deep fields D1-4. Error bars were determined from Gaussian Statistics. Also shown in filled black circles are galaxy number counts for the average of our four fields , for which error bars represent the root-mean-square deviation. Both open red symbols and black filled circles were shifted horizontally from the midpoint of the bin for clarity.}
   \label{fig:ncpe}
\end{figure}

\subsection{Passive \zs2 galaxies}\label{sec:PE_numbercounts}

Galaxy number counts for $z\sim2$ passive galaxies are shown in Figure \ref{fig:ncpe}.  As before, we compare our results with those of several different authors that used the \bzk selection technique. For Hartley et al. (2008) and McCracken et al. (2010), we included a shaded areas that represents their upper and lower limits. For Hartley et al. (2008) the upper limit represents the possibility that all of their non-detections in B and z' were in fact $pBzK$ galaxies. Although their lower limit exhibits a turnover in the number counts at faint magnitudes, once the upper limits is considered, there is no evidence for turnover in their number counts before their $K_{s}=23$ limit.

For McCracken et al. (2010), the lower limit of their shaded region indicates galaxies that are unambiguously $pBzK$, while the upper limit also includes galaxies with ambiguous SF/PE classifications due to their lack of $B$-band detection. Even taking into account their worst case scenario, they observe a flattening of the number counts although within the limits the data are consistent with no turnover.

Our results seem to be in good agreement with most of the other studies, especially with McCracken et al. (2010).  The only significant discrepancy is with Hartley et al. (2008). McCracken et al. (2010) suggested that the colour transformation used by Hartley et al. (2008) did not reproduce the classic $BzK$ selection well, leading to discrepancies between these authors and other studies.  We again do not show the results of Bielby et al.\ (2012) since their use of the classic three-filter $BzK$ selection with the relatively shallow CFHTLS $g$-band data can lead to significant biases in their selection at fainter magnitudes.

\subsection{Detailed comparisons}\label{sec:detailed_numbercounts}

\begin{figure}
   \centering
   \includegraphics[width=7.5cm]{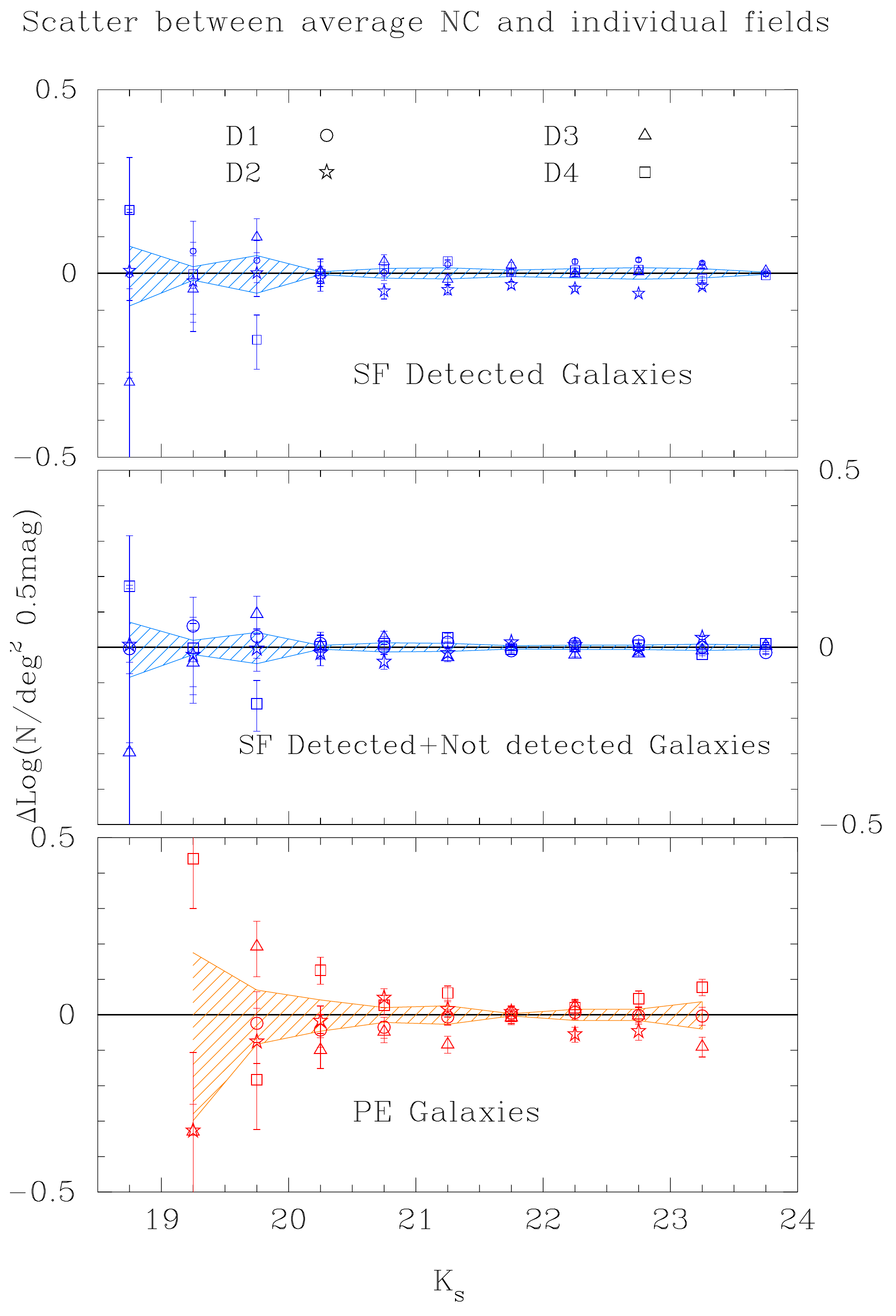} 
   \caption{Scatter between our different populations and their respective average number counts. The top panel shows the scatter for only galaxies that were classified as star-forming and detected in g, while the middle panel also shows star-forming galaxies regardless of whether or not were detected in g. The bottom panel shows the scatter for passively evolving galaxies.}
   \label{fig:scatterfields}
\end{figure}

\begin{figure}
   \centering
   \includegraphics[width=7.5cm]{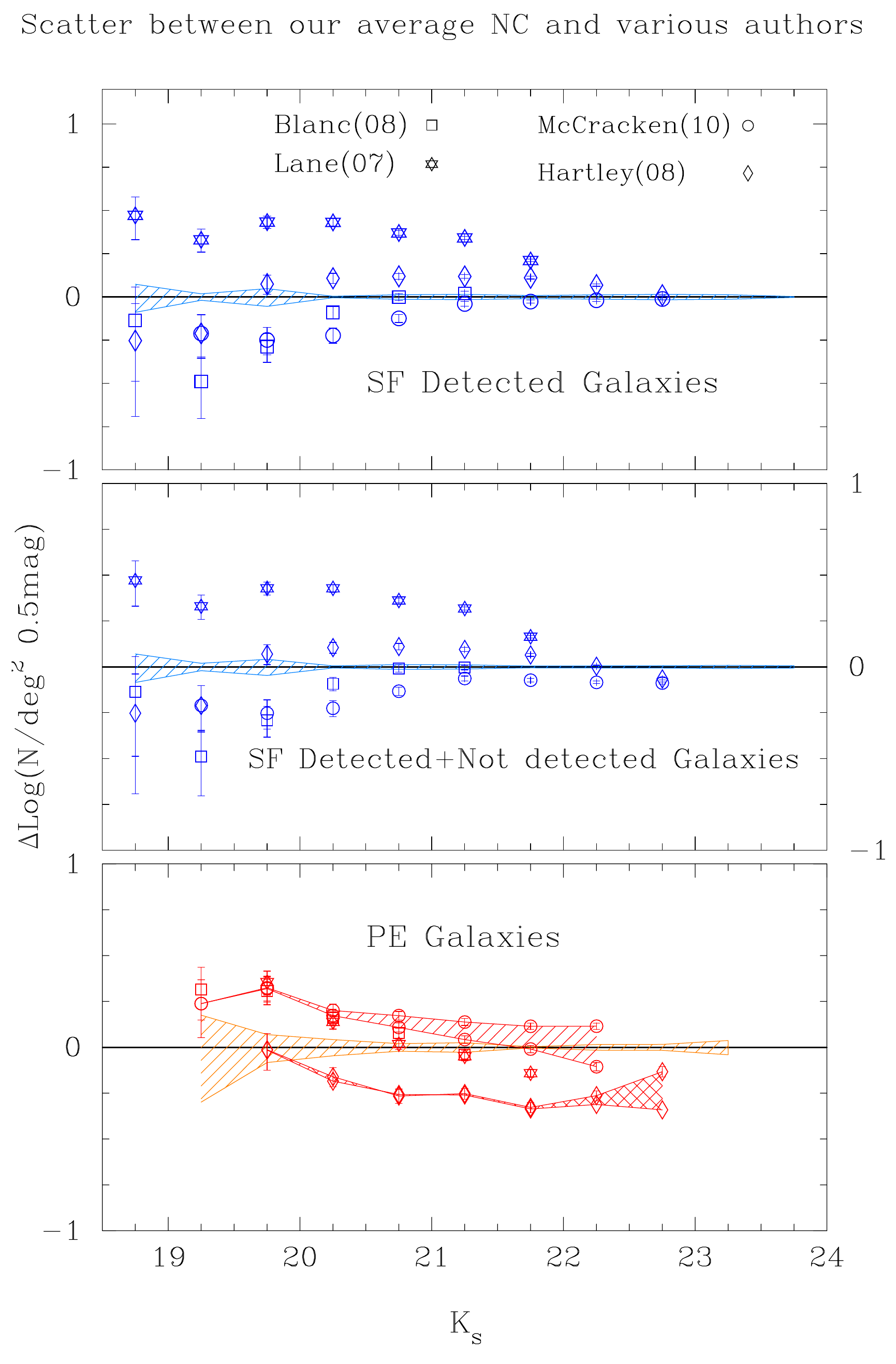} 
   \caption{Scatter between our average number counts and those of various authors. The top panel shows the scatter for only galaxies that were classified as star-forming and detected in g, while the middle panel also shows star-forming galaxies regardless of whether or not were detected ing. The bottom panel shows the scatter for passively evolving galaxies. Following the representation given in Figure \ref{fig:ncpe}, the shaded areas represent the upper and lower limits (see text) of Hartley et al. 2008 (Cross-hatched region) and McCracken et al. 2010 (Hatched region).}
   \label{fig:scatter}
\end{figure}

In \S~\ref{sec:all_numbercounts}--\ref{sec:PE_numbercounts} we found that our results are in good general agreement with those of most other recent surveys. Our study's large total area, split over four independent sightlines, allows us to carry out much more detailed comparisons, in particular focusing on the issue of cosmic variance in addition to purely systematic variations between different surveys. To do so we will take the average counts in our four fields as baseline and then ask (i) how much field-to-field variation there is within our data?, and then (ii) how big are the differences between different studies? 
 
 Figure \ref{fig:scatterfields} shows the residuals of our number counts after subtracting off the mean counts computed by averaging the results of the four fields. The shaded regions represent the uncertainty in the mean computed by summing the individual fields' uncertainties in quadrature. Except at the very brightest magnitudes, which are dominated by small number statistics, the field-to-field variance is quite small, though significant when compared to the errorbars. In particular, the D2 (COSMOS) field shows signs of a deficit of \zs2 galaxies compared to our four-field average. Most significantly, here, the number of faint (\Ks$>$22) passive galaxies is lower than the average by about 10\%.  The number of $g$-detected star-forming galaxies is also low in D2, though the discrepancy for the SF object disappears once $g$-undetected galaxies are included. There are other departures from the average counts --- e.g., the D4 field has a surplus of PE galaxies --- but the deficits in D2 (COSMOS) are of particular note given the importance of this field in extragalactic studies (Scoville et al.\ 2007). We will touch on this issue again in \S~\ref{sec:LF} when we discuss the \zs2 luminosity functions. 
 
The observed field-to-field variance can be regarded as either due to large-scale structure present in the four fields (``cosmic variance"), or systematic experimental variations such as photometric calibration errors etc.  Because any inter-field differences cannot be smaller than that due to cosmic variance, field-to-field variance we observe puts an upper limit on the amount of cosmic variance present on the scales $\sim$0.5 deg$^2$. 

Turning to variance between studies, in Fig.~\ref{fig:scatter} we plot the residuals of various surveys' number counts after subtracting off the average of our study's four fields.  As we mentioned before, all results are in broad agreement except for the star-forming galaxy counts of Lane et al. (2007) and the PE galaxies of Hartley et al. (2008). Examining the data more closely we see that there nevertheless exist systematic differences even between the other surveys.  When comparing any two surveys, the  offsets are often magnitude-dependent and can range from zero to as much as $\sim$0.3 dex.  Since cosmic variance is relatively small, as shown by the small field-to-field scatter among the four CFHT fields, the differences between surveys are likely to be due to systematic effects.  In particular, although all surveys use BzK-like selection, they all rely on somewhat different filter sets and --- hence --- transformations to the original Daddi et al.\ (2004) filter set.  It is thus very likely that all these ``BzK" studies select somewhat different, if related, populations of galaxies, leading to the observed differences in number counts. 

We conclude that cosmic variance in $\sim$0.5 deg$^2$ \zs2 surveys is relatively small in comparison to systematic calibration effects.

\subsection{Comparison with models}\label{sec:models}

Number counts from some semi-analytic models are also shown in Figures \ref{fig:ncsf} and \ref{fig:ncpe} using magenta lines. 

 Solid magenta lines represent models from Kitzbichler \& White (2007) as presented by McCracken et al. (2010). In both figures the Kitzbichler \& White semi-analytic models over-predict the number of faint galaxies and under-predicts the number of bright galaxies as compared with observational results.  

Semi-analytic models from Merson et al. (2012) are shown as magenta dashed and dot-dashed lines. The dashed line represents model $BzK$ galaxies brighter than $K_{AB}$$\leq$24. Although the number counts of star-forming galaxies show good overall agreement with observations at the bright end ($K_{s}$$\sim$ 20), the model over-predicts the number of faint SF galaxies.

To asses the importance of a B-band depth on BzK selection, Merson et al. (2012) recalculated their predicted number counts of SF-$BzK_{s}$ and PE-$BzK_{s}$ assuming a B-band detection limit of $B_{AB}$=28 in addition to the original K-band limit of $K_{AB}$$\leq$24. These results are shown as a dot-dashed magenta lines in Figures \ref{fig:ncsf} and \ref{fig:ncpe}. As before the predicted number counts are not in full agreement with the data, but the the predictions for PE-$BzK_{s}$ galaxies are somewhat improved.

In light of the discussion in \S~\ref{sec:detailed_numbercounts}, we note that inter-survey differences are likely dominated by systematic uncertainties in, e.g., colour transformations. Such systematics likely also contribute to the disagreement between models and data, but given the size of this disagreement, it remains likely that the models do not yet capture all the relevant physics.


\section{Rest-Frame $R$-Band Luminosity Functions}\label{sec:LF}

In this section we measure and analyze the LF of passive and star-forming galaxies at \zs2. 

\subsection{Estimating the Luminosity Function}

We break the construction of the LF into three steps:  (1) correction of the observed number counts for incompleteness, (2) determination of the survey effective volumes (\Veff), and (3) conversion of apparent $K$-band to absolute $R$-band magnitudes.  The first of these three steps, incompleteness correction, was already discussed in \S~\ref{sec:counts}.  The other two steps are described below. 

For simplicity we assume that our galaxies come from a well-defined redshift interval that can be described by a simple top-hat redshift selection function.  The boundaries of this selection function are chosen based on examining the same models that we used in designing the colour-colour selection regions in \S~\ref{sec:gzhk}.  For concreteness, for star-forming galaxies we adopt the 100~Myr continually star-forming model with \ebv=0.2 and this gives \Veff=1.69$\times10^7$ Mpc$^3$ deg$^{-2}$.  Other reasonable combinations of age and extinction give values that are up to $\sim$1.5 times higher or lower than this fiducial.  For passive galaxies we adopt the 1~Gyr-old single burst model with no dust, which results in \Veff=1.77 $\times10^7$ Mpc$^3$ deg$^{-2}$.  Older models give \Veff\ up to $\sim$1.3 times larger than this fiducial value and younger models give lower \Veff --- as low as $\sim$3 times lower for young ($\sim$$10^{8.5}$ Gyr) instantaneous bursts. 

Next, in the usual way (see, e.g., Lilly et al.\ 1995; Sawicki \& Thompson 2006), absolute $R$-band magnitudes are calculated from the observed $K$-band using 
\begin{eqnarray}\label{eq:KtoM_R}
\label{eq:deltamag}
M_{R}=m_{\lambda_{obs}} - 5 log \left(\frac{D_{L}}{10pc}\right) + 2.5 log (1+z)  \nonumber \\
+  \left(m_{R}-m_{\lambda_{obs}/(1+z)}\right).
\end{eqnarray}
Here $D_{L}$ is the luminosity distance and the last term is the k-correction colour between rest-frame $R$ and observed $K_{s}$ -bands.  We assume $z=2$ for the calculation and note that because of our choice to work at rest-frame $R$ the k-correction colour term will be small and insensitive to the assumed galaxy spectral energy distributions. Assuming here that our galaxies are at $z=2$ may introduce two biases: First, if the median redshift of our sample differs from $z=2$ the overall magnitude calibration of the LF will be also offset.  Second, because our \zs2 samples contain galaxies over a range of redshifts, the luminosity function will suffer a degree of smearing along the magnitude direction. However, these effects are not likely to dramatically affect our luminosity function measurement: taking $z=1.5$ or 2.5 instead of $z=2$ gives a systematic offset of $\sim \pm 0.5$ mag and this is small compared to the features in the luminosity function that we will be discussing in the next section. Furthermore, the assumption that the median redshift of our sample lies at z$\sim$2 has been supported by previous studies (e.g.,  Daddi et al. 2004, Reddy et al. 2005, Hayashi et al. 2007,  P\'erez-Gonz\'alez et al. 2008).

\begin{figure}
   \centering
   \includegraphics[width=7.7cm]{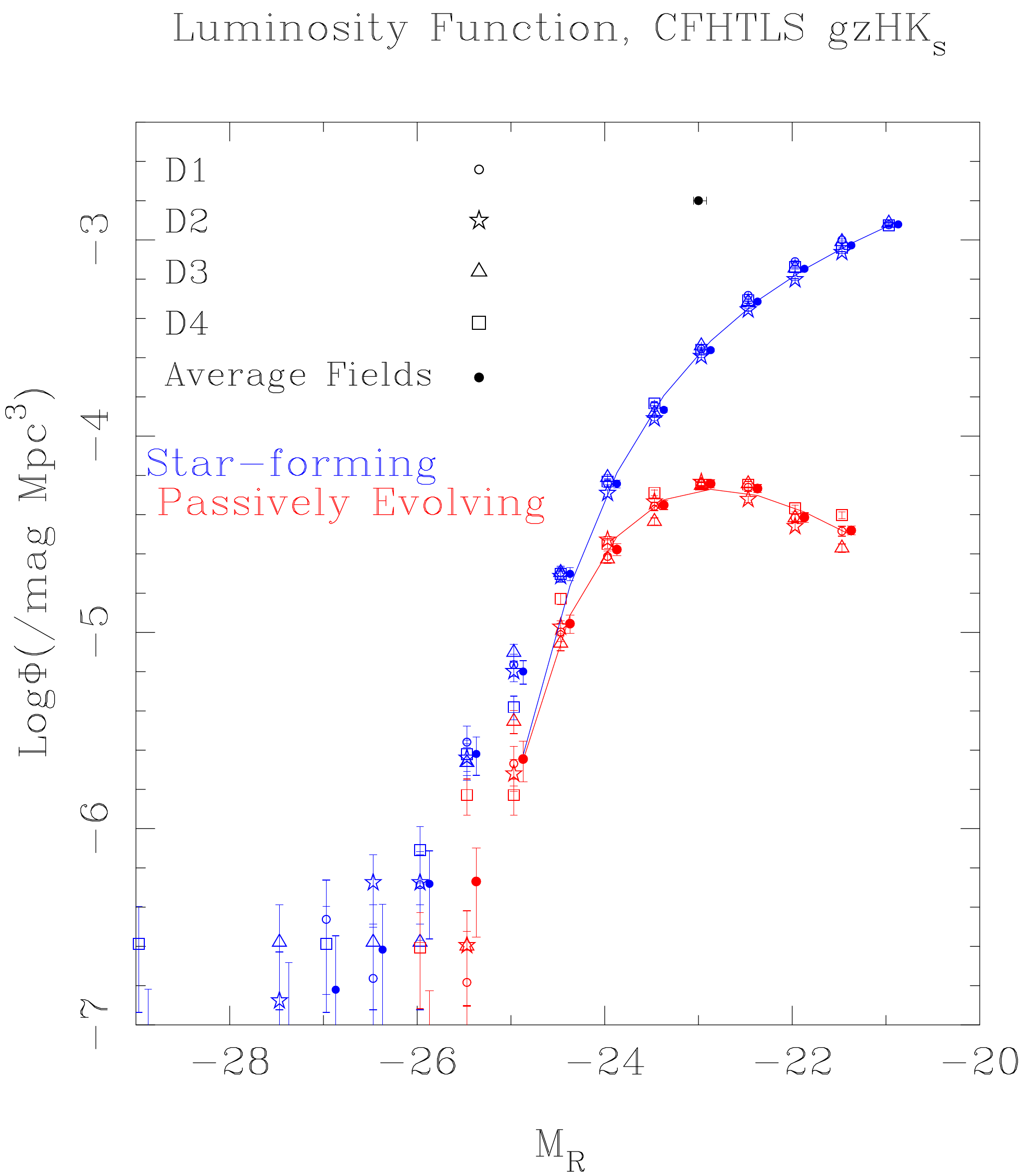} 
   \caption{Rest-frame $R$-band luminosity functions for \zs2 passive (red symbols) and star-forming (blue) galaxies. Results from each Deep Field are represent with a different open symbol, while the averaged results are shown with filled points.  Points have been shifted horizontally from the bin mid-point for clarity.  Error bars for individual fields are $\sqrt{N}$, while the error bars for the average fields represent their sum in quadrature.  The best-fitting Schechter functions are shown  as blue and red solid lines. In the upper part of the figure, the horizontal error bar is an estimate between our best guess in absolute magnitude and the difference between this best guess and our upper and lower limits at $z\sim2$.}
   \label{fig:lf}
\end{figure}

Using the steps described above, we convert the incompleteness-corrected $\phi_{f}(m)$ number count values determined in \S~\ref{sec:counts} to the rest-frame $R$-band luminosity function $\phi_{f}(M_R)$. We plot them in Figure \ref{fig:lf}, where blue is used to identify star-forming galaxy LFs and red represents passively-evolving ones.  

\begin{figure*}
	\begin{minipage}{0.5\linewidth}
\centering
   \includegraphics[width=7.9cm]{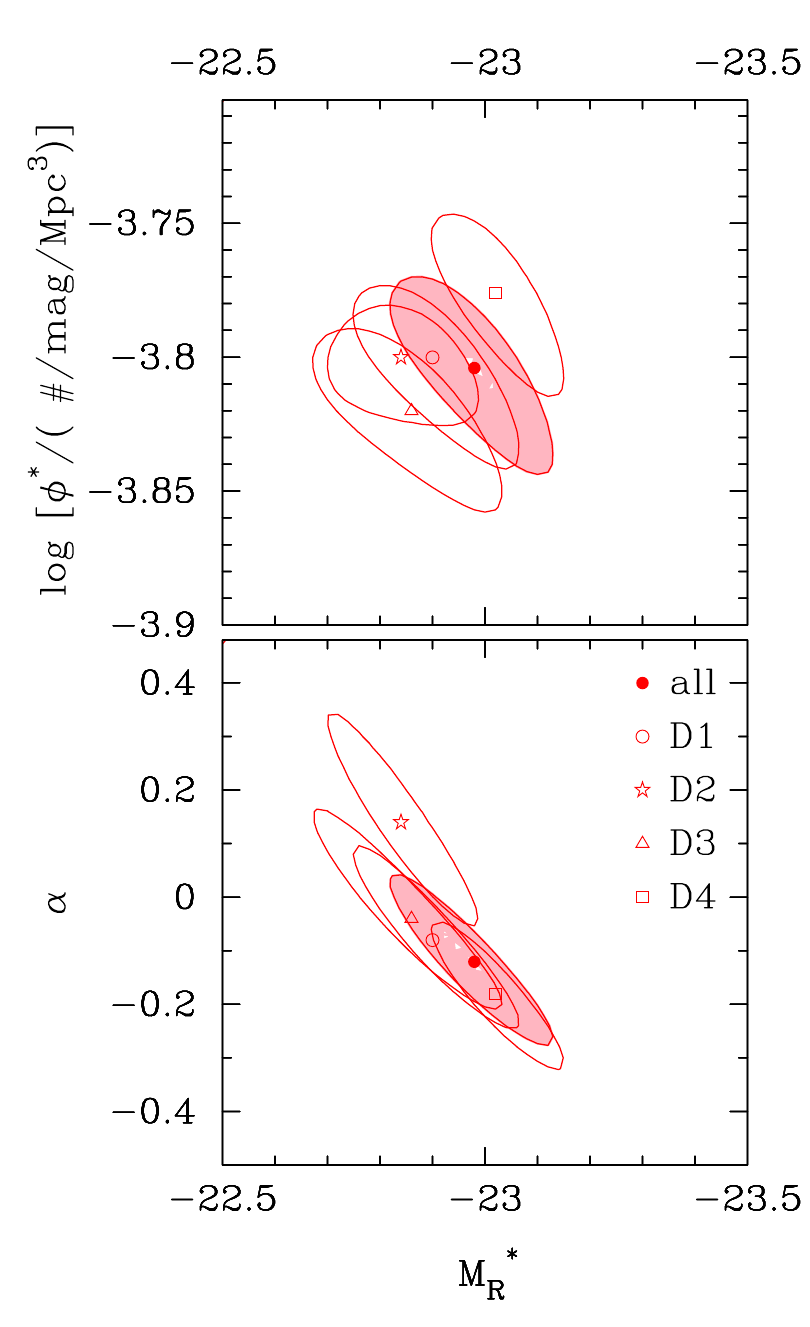} 
\end{minipage}%
\begin{minipage}{0.5\linewidth}
\centering
\includegraphics[width=7.38cm]{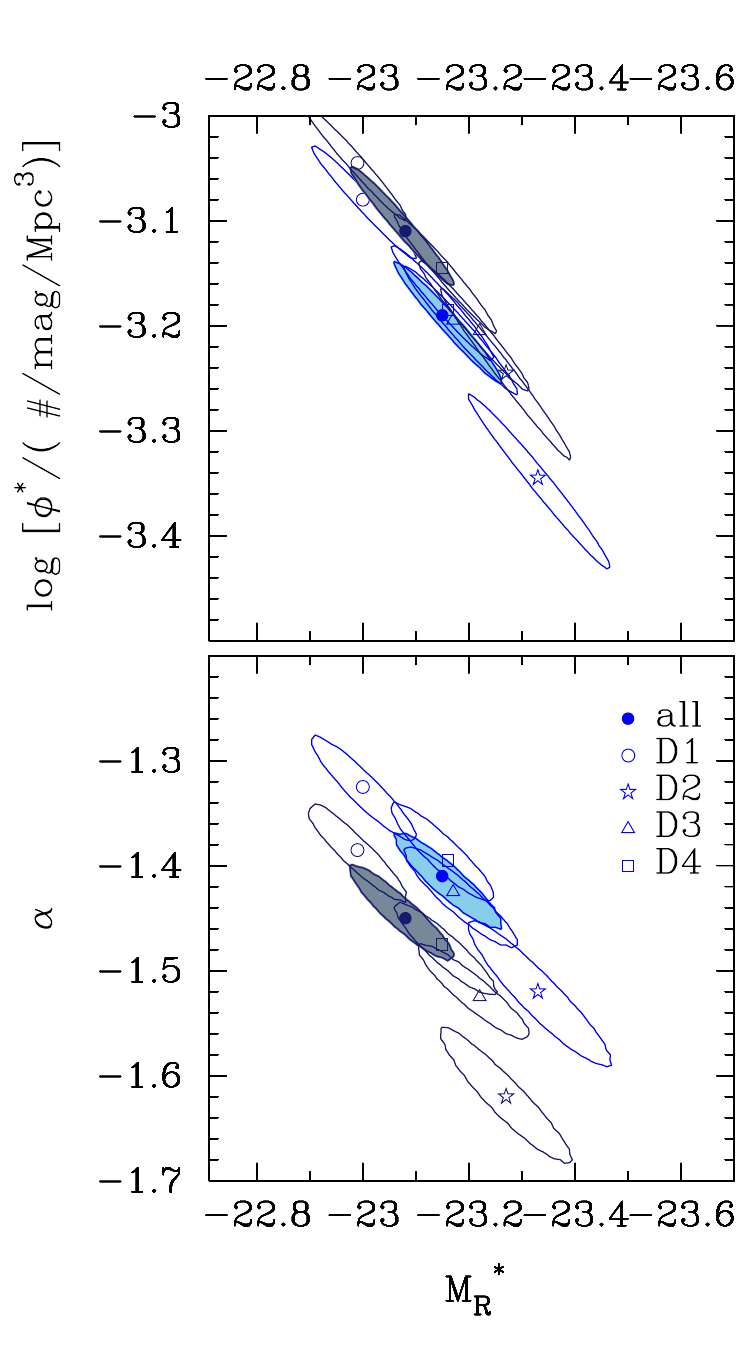}
\end{minipage}
\caption{Confidence regions (1$\sigma$) on the Schechter LF parameters for the passively-evolving (left) and star-forming (right) populations. The filled symbols and shaded contours denote the averages over the four fields.  In the right panels, star-forming galaxies detected in $g$-band are represented in lighter blue while dark blue is for galaxies that were classified as star-forming regardless of whether or not they were detected in $g$-band.
}	\label{fig:schechter}
\end{figure*}

\begin{table*}
	\centering
	\caption{Schechter Function Fits and Luminosity Densities for the four CFHTLS/WIRDS Deep Fields. \emph{Note:} $\rho_{L}$ is at rest-frame $R$ and is given in logarithmic units of erg $s^{-1}$ $Hz^{-1}$ $Mpc^{-3}$}
		\begin{tabular}{|c|clc|c|c|c|}
		\hline
		\bf{Population}&\bf{Field}&$M^{*}$&\bf{$Log_{10}$$\phi^{*}$ $[Mpc^{-3}]$}&\bf{$\alpha$}&$\rho_{L}$\\\hline\vspace{0.1cm}
		
		\bf{$g$-detected SF}&Average&-$23.15^{+0.09}_{-0.11}$&-$3.19^{+0.05}_{-0.06}$&-$1.41^{+0.05}_{-0.04}$&$26.90^{+0.07}_{-0.08}$\\\vspace{0.1cm}
		
		&D1&-$23.00^{+0.09}_{-0.01}$&-$3.08^{+0.05}_{-0.05}$&-$1.32^{+0.04}_{-0.05}$&$26.89^{+0.06}_{-0.07}$\\\vspace{0.1cm}
		
		&D2&-$23.33^{+0.13}_{-0.13}$&-$3.34^{+0.08}_{-0.09}$&-$1.52^{+0.07}_{-0.07}$&$26.90^{+0.11}_{-0.13}$\\\vspace{0.1cm}
		
		&D3&-$23.17^{+0.09}_{-0.12}$& -$3.19^{+0.05}_{-0.07}$&-$1.42^{+0.04}_{-0.05}$&$26.91^{+0.07}_{-0.09}$\\\vspace{0.1cm}
		
		&D4&-$23.16^{+0.1}_{-0.08}$&-$3.18^{+0.06}_{-0.05}$&-$1.39^{+0.05}_{-0.04}$&$26.90^{+0.07}_{-0.07}$\\\hline\vspace{0.1cm}

		\bf{$g$-det \& not detected SF}&Average&-$23.08^{+0.10}_{-0.09}$&-$3.11^{+0.06}_{-0.05}$&-$1.45^{+0.05}_{-0.04}$&$26.98^{+0.08}_{-0.07}$\\\vspace{0.1cm}
		
		&D1&-$22.99^{+0.09}_{-0.09}$&-$3.04^{+0.04}_{-0.06}$&-$1.38^{+0.04}_{-0.05}$&$26.96^{+0.06}_{-0.08}$\\\vspace{0.1cm}
		
		&D2&-$23.27^{+0.12}_{-0.12}$&-$3.24^{+0.08}_{-0.08}$&-$1.62^{+0.07}_{-0.06}$&$27.08^{+0.11}_{-0.13}$\\\vspace{0.1cm}
		
		&D3&-$23.22^{+0.11}_{-0.09}$&-$3.20^{+0.06}_{-0.06}$&-$1.52^{+0.05}_{-0.04}$&$27.00^{+0.08}_{-0.09}$\\\vspace{0.1cm}
		
		&D4&-$23.15^{+0.09}_{-0.10}$&-$3.14^{+0.04}_{-0.06}$&-$1.47^{+0.04}_{-0.05}$&$26.99^{+0.06}_{-0.08}$\\\hline\vspace{0.1cm}

		\bf{Passive}&Average&-$22.98^{+0.16}_{-0.14}$&-$3.80^{+0.03}_{-0.04}$&-$0.12^{+0.16}_{-0.14}$&$26.02^{+0.06}_{-0.08}$\\\vspace{0.1cm}
		
		&D1&-$22.90^{+0.14}_{-0.16}$&-$3.80^{+0.03}_{-0.04}$&-$0.08^{+0.16}_{-0.16}$&$25.99^{+0.07}_{-0.06}$\\\vspace{0.1cm}
		
		&D2&-$22.84^{+0.12}_{-0.14}$&-$3.80^{+0.02}_{-0.02}$&$0.14^{+0.20}_{-0.17}$&$26.01^{+0.07}_{-0.07}$\\\vspace{0.1cm}
		
		&D3&-$22.86^{+0.18}_{-0.16}$&-$3.82^{+0.03}_{-0.04}$&-$0.04^{+0.20}_{-0.16}$&$25.96^{+0.08}_{-0.09}$\\\vspace{0.1cm}
		
		&D4&-$23.02^{+0.12}_{-0.12}$&-$3.77^{+0.03}_{-0.04}$&-$0.18^{+0.12}_{-0.14}$&$26.06^{+0.06}_{-0.07}$\\
		
		\hline
		\end{tabular}
	   	\label{tab:schechter}
\end{table*}

\subsection{Analysis of the Luminosity Functions}

The luminosity function data are fit using the Schechter (1976) LF parametrization, 
\begin{eqnarray}
\label{eq:phimodel}
\phi_{model}=\phi^{*} 0.4 \ln(10) \{{\rm dex} \left[0.4(M^{*}-M)\right]\}^{(\alpha+1)}\nonumber \\
\times  \exp \left(-10^{0.4(M^{*}-M)}\right), 
\end{eqnarray}
where only points fainter than $M_R = -25$ were used in the fit. The best-fitting functions are plotted as red and blue curves in Fig.~\ref{fig:lf}, the best-fitting Schechter parameter values are given in Table \ref{tab:schechter}, and Figure~\ref{fig:schechter} shows the error contours for the Schechter parameters for the star-forming and passive populations.  

The faint-end slope for SF galaxies is $\alpha= -1.41^{+0.05}_{-0.04}$, or $\alpha=-1.45^{+0.05}_{-0.04}$ if $g$-non-detections are also included in the fit.  Combining these two results gives $\alpha = -1.43\pm0.02^{+0.05}_{-0.04}$ (systematic and random uncertainties, respectively).  It is worth noting that our SF galaxy faint-end slope is in very good agreement with the BX galaxy $\alpha=-1.47^{+0.24}_{-0.21}$ of Sawicki (2012b). While the BX and SF-BzK populations are necessarily somewhat different, both these techniques aim to select star-forming galaxies at \zs2 and do give significant overlap (e.g., Reddy et al.\ 2005). The similarity of the observed BX and SF-BzK faint-end slopes, while not surprising, is reassuring. 

Turning to passively-evolving galaxies, our PE galaxy faint-end slope $\alpha=-0.12^{+0.16}_{-0.14}$ is in good agreement with the determinations of the low-mass end of the \zs2 PE galaxy stellar mass functions (e.g., Kajisawa et al.\ 2011, Wuyts et al.\ 2008), but our $\alpha$ measurement is much better constrained than in those earlier studies, owing to our large $\sim$5000-object sample of \zs2 passive galaxies.  It is now clear that the numbers in the faint/low-mass end of the PE-BzK galaxy population at \zs2 trend strongly downwards, with a steeply declining slope as is also seen for passive galaxies at lower redshifts.  The most common passive galaxy at \zs2 has $M_R\sim -23$, an issue that we will revisit in \S~\ref{sec:SMF} where we examine galaxy stellar masses.

With the Schechter fit parameters in hand, integrating the Schechter function gives the sum of all the light from all galaxies in  a unit of volume:
\begin{equation}
\rho_{L}=\int^{\infty}_{0}\hspace{0.1cm} L\phi\left(L\right)dL =\phi^{*}L^{*}\Gamma\left(\alpha+2\right),
\label{eq:lumden}
\end{equation}
where $\Gamma(x) = \int^{\infty}_{0}t^{x-1}e^{-t}dt$ is the usual gamma function.  Column 6 of Table \ref{tab:schechter} lists the rest-frame $R$-band luminosity densities in our four fields. The integrated luminosity density does not vary significantly from field to field.  Typical departures from the averaged-field values are $\lesssim$10\%, which is entirely within the error budget. However, these luminosity density values hide underlying differences between fields in the shapes of the LFs.  As with the number counts (\S~\ref{sec:detailed_numbercounts}), there are field-to-field differences in the LF shapes (Fig.~\ref{fig:lf}) and these differences are also reflected in differences in the Schechter function parameters (Fig.~\ref{fig:schechter}). In particular, the D2 (COSMOS) field is especially anomalous: there is a deficit of faint PE galaxies in this field compared to the average, and this is reflected in a shallower value of the faint-end slope parameter $\alpha$ (see bottom left panel of Fig.~\ref{fig:schechter}). Similarly, the LF of SF galaxies in the D2 field also appears significantly different from the average LF and these differences would be increased if we were to exclude the potentially anomalous D2 field from that average.  The observed field-to-field variation is not large --- the deficit of faint galaxies in D2 is $\sim$10\% compared to the average --- and is  likely consistent with fluctuations due to large scale structures. However, the fact that the D2 (COSMOS) field is anomalous is notable because of the importance of this extremely well-studied field in galaxy formation studies.

Our luminosity function for passive galaxies shows a turnover (rather than just a flattening), with the peak at $M_R \sim -23$. This implies that the passive galaxy population at \zs2 is dominated by bright (massive) objects, a phenomenon that is consistent with the ``downsizing" scenario. Downsizing, as introduced by Cowie et al. (1996), implies that star-formation stops in massive systems first and in lower-mass systems later. Our observations are consistent with the downsizing scenario because if we interpret rest-frame $R$-band luminosity as a surrogate for the stellar mass of a galaxy, then at the low mass end most galaxies are still forming stars, while at the massive end passive and star-forming galaxies are present in roughly equal numbers. 
It is important to note that both a turnover and a simple flattening in the luminosity function are consistent with downsizing. However, a turnover --- which is now observed in the data --- represents a much more interesting scenario because it suggests that there is a characteristic luminosity (i.e., mass) at which galaxies are most likely to evolve into passive systems. We examine this issue further in the next section.

\section{Stellar Mass Functions}\label{sec:SMF}

\begin{figure*}\label{fig:SMF}
	\begin{minipage}{0.5\linewidth}
\centering
   \includegraphics[width=7.5cm]{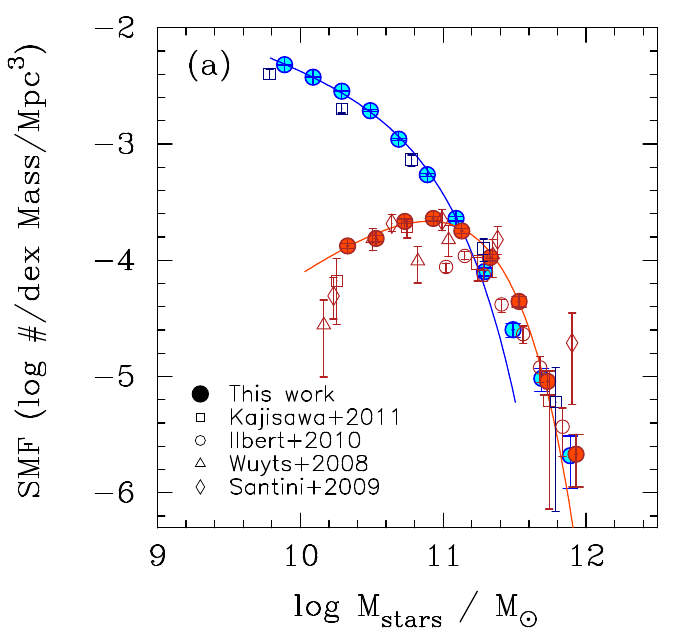} 
\end{minipage}%
\begin{minipage}{0.5\linewidth}
\centering
\includegraphics[width=7.8cm]{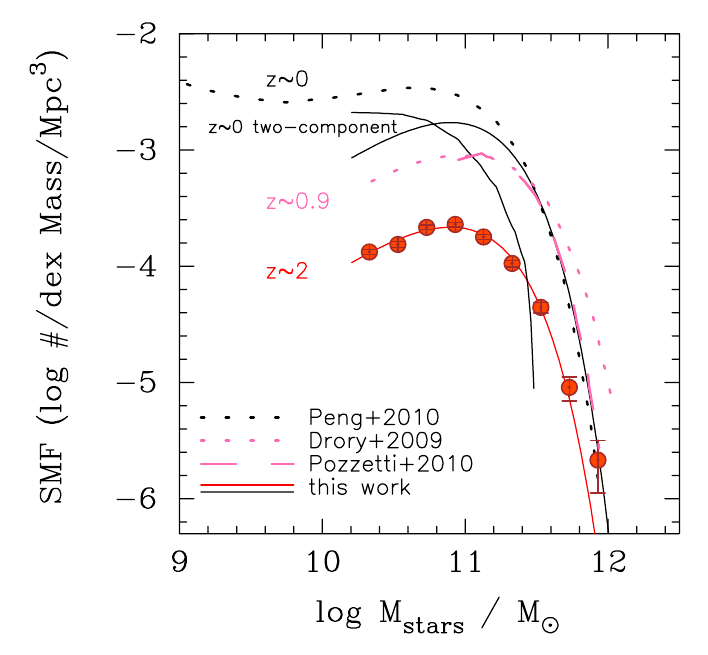}
\end{minipage}
\caption{Stellar mass functions.  The {\emph{left panel}} shows the \zs2 stellar mass functions by galaxy type and in comparison with some previous studies. Some of the data points have been slight offset horizontally for clarity. The most common PE galaxy at \zs2 has a stellar mass of $\sim 10^{11}$\Msun\ (Salpeter IMF).   The {\emph{right panel}} shows the time evolution of the mass functions of PE galaxies. The two solid black curves, labelled ``\zs0 two-component", represent the observed \zs2 SMF scaled up uniformly by a factor of 8 and the residual of subtracting that scaled \zs2 SMF from the observed \zs0 SMF. It thus appears plausible that a population of massive PE galaxies builds up uniformly in number, starting already before \zs2 and continuing to the present day. A time-invariant peak mass of $\sim 10^{11}$\Msun\ is a hallmark of this build-up.
}	\label{fig:SMF}
\end{figure*}

Rest-frame $R$-band light in galaxies is generated in large measure by the long-lived low-mass stars that in stellar populations originating from reasonable IMFs can be expected to contain most of a galaxy's stellar mass.  Via a simple light-to-mass conversion, rest-frame $R$-band luminosities can thus be used to estimate stellar masses of distant galaxies, and, consequently, rest-frame $R$-band LFs can be used as a proxy for stellar mass functions (SMFs).  

Using this approach, in this section we make a straightforward estimate of the \zs2 SMFs from the LFs determined in \S~\ref{sec:LF}.  The conversion from $R$-band luminosity is more straightforward for passively evolving objects that it is for star-forming ones, and so we focus our analysis on the PE SMF, only briefly touching on the SF SMF.

\subsection{From LF to SMF}

We apply a very simple procedure to convert from the LFs we determined in \S~\ref{sec:LF} to stellar mass functions, treating the PE and SF conversions separately. 

\subsubsection{Passively evolving galaxies}

For the PE galaxy population we derive the light-to-mass conversion based on the luminosity of a $z=2$, 700~Myr-old dust-free single stellar population from the Bruzual \& Charlot (2003) model library.  This simple model is consistent with the definition of PE \bzk and \gzhk galaxies (Daddi et al.\ 2004 and also \S~\ref{sec:gzhk}).  In performing stellar mass estimates it is important to account for stellar mass returned to the ISM by mechanisms such as stellar winds and supernova explosions, and for our 700~Myr-old single stellar population this returned fraction is 23\% (see the models of Bruzual \& Charlot, 2003).  This light-to-mass conversion is sensitive to the assumed age of the stellar population --- for example, a 2~Gyr-old single stellar population would give a stellar mass that is $\sim2\times$ heavier than what we assume; however, such a 2~Gyr stellar population, observed at $z=2$ would have to have formed at \zs 4.8 and given the rapid, factor of \s10 growth in the stellar mass density of the universe between \zs5 and \zs2 (e.g., P\'erez-Gonz\'alez et al.\ 2008, Elsner et al.\ 2008, Yabe et al.\ 2009, Sawicki 2012a), it seems unlikely that many objects in our population of the \zs2 PE galaxies would have been already quiescent as early as \zs5. Conversely, galaxies with stellar populations significantly younger than our assumed 700~Myr cannot be considered to be truly quiescent yet by our definition, and so we settle on 700~Myr as a reasonable age that is also consistent with the classic definition of \bzk galaxies (Daddi et al.\ 2004).

Bearing in mind caveats regarding the simplicity of our light-to-mass conversion, we convert our observed \zs2 PE galaxy LF to a PE galaxy SMF. We show the result for our averaged fields (filled red points) in the left panel of Fig.~\ref{fig:SMF}, along with our Schechter function fit to the data (red curve).  The PE galaxy SMF at \zs2 has a steeply rising massive end, and a declining low-mass end.  The peak of the SMF, which corresponds to the mass of the most common PE galaxy at \zs2, is at \s10$^{11}$\Msun.  Open red symbols in Fig.~\ref{fig:SMF} show previous determinations of the PE SMF at \zs2 (Wuyts et al.\ 2008, Santini et al.\ 2009, Ilbert et al.\ 2010, Kajisawa et al.\ 2011) which we transformed, when necessary, to our adopted Salpeter (1955) IMF. Our PE galaxy SMF agrees well with these earlier determinations of the PE MF, but our sample contains many more objects and thus has much better statistics than these earlier studies. 

\subsubsection{Star-forming galaxies}
	
The light-to-mass conversion for SF galaxies is more uncertain than that for PE systems because of the uncertainties about their star formation histories (we assume constant SFR but in practice SFHs could be decreasing, increasing, or variable with time) and interstellar extinction (we assume constant extinction whereas in practice extinction at \zs2 is a function of galaxy luminosity --- Sawicki 2012b). Despite these issues we proceed by adopting the 200~Myr-old, constant star formation Bruzual \& Charlot (2003) stellar population with $E(B-V) = 0.3$.   Our resulting SF galaxy SMF, shown with filled blue points in Fig,~\ref{fig:SMF}, is a reasonable match to that determined for \zs 2 Kajisawa et al. (2011) using full multi-wavelength SED-fitting applied to the deep observations of the GOODS-N region.  However, because of the higher systematic uncertainties associated with the SF galaxy SMF, in the next section we focus primarily on discussing the SMF of the PE systems, touching on SF galaxies only briefly at the end.

\subsection{The non-evolving shape of the PE stellar mass function}

In the right panel of Fig.~\ref{fig:SMF} we compare our \zs2 PE galaxy SMF with SMFs observed at lower redshifts.  The two \zs0.9 SMFs are both from data in the COSMOS field, one derived using a spectroscopic sample (Pozzetti et al. 2010), the other using a large photometric redshift one (Drory et al.\ 2009); the local PE galaxy SMF is from the analysis of SDSS data by Peng et al.\ (2010).  All are plotted on a common Salpeter IMF scale. 

The shape of the \zs2 PE galaxy SMF is remarkably similar to that at \zs0.9.
In common with \zs0.9 observations, the \zs2 has a steeply rising massive end, a peak around $10^{11}$\Msun, and a declining low-mass end, although the overall normalization in number density is \s4$\times$ higher at \zs0.9 than it is at \zs2. 
This remarkable similarity of the shape between \zs2 and \zs0.9, with just a shift of a factor-of-four in normalization across all masses, suggests that the same physical process may be responsible for forming the PE galaxy SMF that's seen at \zs2 as that at \zs0.9.  In this scenario the population of PE galaxies grows at a rate that gives a uniform increase of a factor of four in number, independent of mass. 

The \zs0 PE galaxy SMF has been measured by numerous authors using SDSS data (e.g., Bell et al.\ 2003, Baldry et al.\ 2008, Peng et al.\ 2010) and in Fig.~\ref{fig:SMF} we use the black curve to show the Peng et al.\ (2010) result.  The overall shape of the \zs0 PE galaxy SMF is different from that at the earlier epochs: while the steep massive end is also present at \zs0, the decline in number density towards lower masses is not as steep as at higher redshifts and there is also a reversal and a raise at very low masses (below $\sim10^{9.7}$\Msun).   Nevertheless, the \zs0 SMF shows a local maximum at $\sim10^{10.6}$\Msun, and this peak mass is similar to the mass of the $\sim10^{11}$ peak at earlier times. 

As mentioned earlier, it appears that evolution of the population of massive galaxies is consistent with simple number density increase between at least \zs2 and \zs0.9.  It is then interesting to consider whether this simple number density growth of massive galaxies may also continue to \zs0. We do this by subtracting a high-mass component from the observed SDSS \zs0 SMF that's identical in shape to the \zs2 SMF and only differers from it in normalization.  Doing so is similar to fitting two Schechter functions to the \zs0 SMF, but rather than allowing all the parameters to vary, here we fix the the shape (i.e., $M^*_{stars}$ and $\alpha$) of the high-mass Schechter function, changing only its $\phi^*$. The result is shown in Fig.~\ref{fig:SMF} using the two black curves, one of which is the scaled-up \zs2 SMF and the other the residual of its subtraction from the SDSS data.  Here we fixed the \zs0 normalization of the \zs2 SMF at $8\times$ that observed at \zs2, although --- given the extreme steepness of the high-mass end --- a range of scaling factors results in reasonable results. It is clear from Fig.~\ref{fig:SMF} that a high-mass component identical (except for its numerical normalization) to that seen at \zs2 can be present in the \zs0 SMF, leaving a second, low-mass Schechter component not seen at higher redshifts. It thus seems possible that a universal mechanism for the formation of high-mass PE galaxies is in action over a large range of redshifts, producing an ever-growing, but self-similar PE galaxy SMF by simply adding new PE galaxies at a rate that only depends on their stellar mass.  Direct evidence for this lies in the similarity between the PE galaxy SMFs at \zs2 and \zs0.9, while the possibility of the \zs0 SMF containing a scaled-up high-mass component extends this argument to the present epoch. 

The SMF growth process described above is consistent with the mass-quenching scenario of Peng et al.\ (2010).  In this mass-quenching model the shut-down of star formation in massive galaxies turns star-forming galaxies into quiescent ones which then ``precipitate" onto the PE galaxy SMF at a rate that changes with time but does not vary strongly with galaxy mass.  Peng et al.\ show evidence for this scenario between \zs0 and \zs1 (when the Universe was about half of its present age). Our data show that if such a process is indeed in operation then it has already left its imprint at \zs2 (\s10~Gyr ago, when the universe was only a quarter of its present age). Furthermore, given that the PE galaxy SMF already has its characteristic shape at \zs2, the process must have already been in operation at even earlier epochs. 

In the Peng et al.\ (2010) model quiescent galaxies form from star-forming ones by two mechanisms: mass quenching and environmental quenching.  In their analysis environmental quenching is not expected to produce significant numbers of PE galaxies until low redshift.  Consequently, unlike at low redshifts, where both mechanisms have contributed to the PE population, at high redshifts we can expect to be observing the pure population of mass-quenched galaxies, uncontaminated by environmentally-quenched objects.  This indeed appears to be the case if we interpret the single Schechter function shape of the \zs2 (and \zs0.9) PE galaxy SMF as due to the single process of mass quenching.  The dominance of mass quenching at \zs0.9--2 (and likely earlier) opens the intriguing possibility of studying this quenching mechanism in isolation from other effects. 

\begin{figure}
   \centering
   \includegraphics[width=7.7cm]{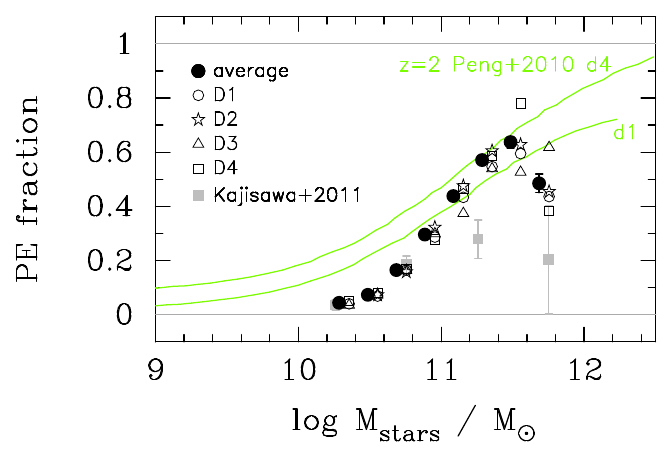} 
   \caption{The fraction of galaxies that are passively evolving at \zs2.  Some of the CFHT points have been offset for clarity. The green curves show the \zs2 models of Peng et al.\ (2010) for the densest (d4) and least dense (d1) quartiles.}
   \label{fig:PEfraction}
\end{figure}

Peng at al.\ (2010) illustrate their proposed quenching scenario with a model of the fraction of galaxies that are passive (the ``PE fraction"). In Fig.~\ref{fig:PEfraction} we reproduce their PE fraction curves for \zs2 where, unlike at low redshift, mass quenching is expected to be the sole important quenching mechanism.  One consequence of the dominance of mass quenching over environmental quenching at high redshift is that the PE fraction is expected to be relatively insensitive to environment. Consequently, unlike at lower redshifts, their models for \zs2 are relatively insensitive to the local density of galaxies --- shown here for their lowest (d1) and highest (d4) density quartiles --- and we can compare the models with our data without concern for environment, which would be difficult to characterize given our lack of spectroscopic redshifts. 

Our data, shown with black points in Fig.~\ref{fig:PEfraction}, are calculated by dividing the PE galaxy SMF by the sum of the PE and SF galaxy SMFs.  The data show a trend that is qualitatively similar to the models:  a low PE fraction at low masses, rising steeply to higher masses.  A similar trend is seen in the data of Kajisawa et al. (2011), though with larger uncertainties and a shallower rise at high masses compared to that observed by us.  The agreement between the data and the models shown here is only qualitative.  However, it should be kept in mind that a range of effects may contribute to the discrepancy, such as differences in definitions of PE and SF galaxies between the data and the models and the systematic uncertainties in our light-to-mass conversions, particularly for the SF galaxies.  We conclude that, qualitatively at least if not yet in detail, at the \zs2 epoch where mass quenching is expected to dominate the PE population, the observed PE fractions are in agreement with the scenario of Peng et al.\ (2010). 

\section{Summary and Conclusions}

The popular \bzk technique (Daddi et al.\ 2004) gives the most complete way to select and classify galaxies at $z\sim2$ that is based only on colour-colour cuts from three different broad-band filters. It provides the opportunity to study galaxy evolution in two populations distinguished by their modes of star formation: currently star-forming as well as passive and hence star-forming in the past. Our adaptation of this technique, combined with the large effective areas (0.4--0.9 $deg^{2}$ each for a total of 2.5 $deg^{2}$) and independent lines of sight of the four Deep Fields in the Canada France Hawaii Telescope Legacy Survey, allowed us to construct and study a large sample of star-forming and passive galaxies in four causally-independent regions of the \zs2 universe.  

In our work, we first adapted the \bzk technique to the CFHTLS+WIRDS \gzk filter set and used it to select $\sim$40,000 \zs2 galaxies brighter than $K_{AB}=24$. Because the CFHTLS $g$-band data are not deep enough to reliably classify the faint galaxies in this \zs2 sample, we then applied a second colour criterion which uses the $H$ filter together with the \gzk\ to classify our \zs2 galaxies into star-forming and passive objects. This \gzhk classification is entirely analogous to that given by the \bzk criteria, allowing us to obtain a sample of $\sim$5000 \zs2 galaxies (to $K_{AB}=23$) that is selected using essentially the popular \bzk  approach.  This forms the largest to date sample of galaxies assembled and classified using $BzK$-like selection and allows the most precise measurement of their number counts, rest-frame $R$-band luminosity functions LFs, and 
stellar mass functions. 

Based on the analysis of these data, our main findings are:

\begin{enumerate}

\item {\emph {Number counts and selection effects:}} Our number counts for both star-forming and passive \zs2 galaxies are in good agreement with those of most recent large-area studies that use \bzk-classified samples (Blanc et al.\ 2008, McCracken et al.\ 2010). However, for star-forming galaxies our results are significantly lower than those obtained by Lane et al. (2007), and for passively evolving galaxies they are higher than those of Hartley et al. (2008). Because the differences between our four large fields are relatively minor (see below), these discrepancies are unlikely to be due to cosmic variance unless the UKIDSS Ultra Deep Field studied by Hartley et al.\ and Lane et al.\ is anomalous in the extreme.  Instead, the number count differences are more likely due a mismatch between the Hartley et al.\ and Lane et al.\ definitions of passive and star-forming \bzk galaxies compared to the ``standard'' Daddi et al.\ (2004) \bzk system. These differences highlight the fact that classification into passive and star-forming objects is relatively arbitrary and can result in significant discrepancies in results between different studies unless care is taken to achieve consistency in object classification. It is likely that all ``$BzK_s$" studies select somewhat different, if related, populations. Differences between populations selected using diverse methods (colour-colour selection, photometric redshifts, morphologies, etc.) as well as those from simulations or theoretical work, are likely to be even larger. 

\item {\emph{Luminosity functions:}}  The passive galaxy LF exhibits a clear peak at $M_R=-23$ and a declining faint-end slope with $\alpha= -0.12 ^{+0.16}_{-0.14}$.  This measurement presents the clearest evidence to date of a turn-over in the LF of \zs2 passive galaxies selected using a \bzk-like technique.  In contrast to the passive galaxies, the LF of star-forming galaxies is characterized by a steep faint-end slope, $\alpha = -1.43\pm0.02(systematic)^{+0.05}_{-0.04}(random)$. This is similar to the steep faint-end slope seen among the BX-classified \zs2 galaxies (Sawicki 2012b). 

\item {\emph { Cosmic variance:}}  The details of both the number counts and luminosity functions are somewhat sensitive to cosmic variance even in our large, $\sim$0.5 deg$^2$,  fields.  In particular, the D2 field (which is in the popular COSMOS survey area) yields LFs that are the most discrepant from the mean. The differences are not large --- the differences in number counts among our $\sim$0.5--1 $deg^2$ fields are typically $<25\%$ for PE galaxies and even less for SF ones. Nevertheless, we should continue to keep cosmic variance in mind, even when dealing with large fields, particularly as we attempt to move towards more and more precise measurements of various observational quantities.

\item {\emph {Stellar mass functions:}}  Our rest-frame $R$-band luminosity functions sample the low-mass stars that contain most of these galaxies' mass, and so we use them to estimate the \zs2 stellar mass functions.  These SMFs are most robust for PE galaxies (which can be assumed to have little dust and relatively simple star formation histories) and so this is where we focus our analysis. The \zs2 PE galaxy SMF has a steeply-raising high-mass end and a declining low-mass end, with a turnover at $M_{stars} \sim 10^{11}$\Msun, the mass of the most common PE galaxy at \zs2.  This picture is consistent with previous \zs2 studies but has much better statistics. The position of the $M_{stars} \sim 10^{11}$\Msun\ turnover and the overall shape of the PE galaxy SMF at \zs2 are very similar (except for its normalization, which is \s4$\times$ lower) to that at \zs0.9. We also show that it is plausible that the \zs0 PE galaxy SMF may contain a subpopulation similar to that seen at higher redshifts.  This  similarity of the PE galaxy SMF shapes over the \zs2 -- 0.9, and possibly also down to \zs0,  suggests a universal mechanism for the formation of high-mass PE galaxies is in action at a wide range of cosmic epochs.  This mechanism, which could be related to the mass-quenching scenario proposed by Peng et al.\ (2010), appears to have already been in operation {\it before} \zs2 in order to produce the observed PE galaxy SMF by that epoch.

\end{enumerate}

As summarized above, our study suggests the existence of a galaxy quenching mechanism that may be universal in time over at least 75\% of the history of the universe, i.e., from \zs2 (or before) until today.  Our evidence for this mechanism lies in the similarity in the shape of the PE galaxy SMF at \zs2 and \zs0.9, and the possibility that the \zs0 SMF contains a component that is also similar to these higher-$z$ SMFs. 
The interesting question to turn to next is that of the nature of the mechanism responsible for this time-invariant, mass-dependent quenching.

\section*{Acknowledgments}

We thank Taro Sato for letting us use his PSF-matched images and completeness simulations in advance of publication,  and Anneya Golob, Bobby Sorba, and the anonymous referee for comments that improved the quality of this manuscript. Computational facilities for this work were provided by ACEnet, the regional high performance computing consortium for universities in Atlantic Canada. ACEnet is funded by the Canada Foundation for Innovation (CFI), the Atlantic Canada Opportunities Agency (ACOA), and the provinces of Newfoundland and Labrador, Nova Scotia, and New Brunswick. This research was financially supported by funds from the Natural Sciences and Engineering Research Council of Canada (NSERC) and by an ACEnet Fellowship. Parts of the analysis presented here made use of the Perl Data Language (PDL) that has been developed by K.\ Glazebrook, J.\ Brinchmann, J.\ Cerney, C.\ DeForest, D.\ Hunt, T.\ Jenness, T.\ Luka, R.\ Schwebel, and C. Soeller; PDL provides a high-level numerical functionality for the perl scripting language (Glazebrook \& Economou, 1997) and can be obtained from http://pdl.perl.org, while additional calculations were carried out using the online cosmological calculator of Wright (2006).

This work is based on observations obtained with MegaPrime/MegaCam and WIRCam. The former is a joint project of CFHT and CEA/DAPNIA, at the Canada-France-Hawaii Telescope (CFHT) which is operated by the National Research Council (NRC) of Canada, the Institut National des Science de lÕUnivers of the Centre National de la Recherche Scientifique (CNRS) of France, and the University of Hawaii. The latter is a joint project of CFHT, Taiwan, Korea, Canada, France, at the CFHT. This work is based in part on data products produced at TERAPIX and the Canadian Astronomy Data Centre as part of the Canada-France-Hawaii Telescope Legacy Survey, a collaborative project of NRC and CNRS, and by the WIRDS (WIRcam Deep Survey) consortium. This research was supported by a grant from the Agence Nationale de la Recherche ANR-07-BLAN-0228.

\end{document}